\documentclass[fleqn,usenatbib]{mnras}

\usepackage{newtxtext,newtxmath}

\usepackage[T1]{fontenc}

\DeclareRobustCommand{\VAN}[3]{#2}
\let\VANthebibliography\thebibliography
\def\thebibliography{\DeclareRobustCommand{\VAN}[3]{##3}\VANthebibliography}

\usepackage{graphicx}	
\usepackage{amsmath}	
\usepackage{array}
\usepackage[table]{xcolor}

\title[The Surface and Interior Conditions of Temperate Sub-Neptune TOI-270~d]{The Surface and Interior Conditions of Temperate Sub-Neptune TOI-270~d}

\author[F. E. Rigby $\&$ N. Madhusudhan]{
Frances E. Rigby$^{1}$\thanks{E-mail: fer29@ast.cam.ac.uk}
and Nikku Madhusudhan$^{1}$\thanks{Email: nmadhu@ast.cam.ac.uk}
\\
$^{1}$Institute of Astronomy, University of Cambridge, Madingley Road, Cambridge CB3 0HA, UK\\
}

\date{Accepted XXX. Received YYY; in original form ZZZ}

\pubyear{2025}

\begin{document}
\label{firstpage}
\pagerange{\pageref{firstpage}--\pageref{lastpage}}
\maketitle

\begin{abstract}

Sub-Neptune planets, with no analogue in our solar system, provide a wealth of information about exoplanet diversity, formation \& evolution, and habitability. Their robust characterisation requires the coupling of physically informed atmosphere and interior models with precise atmospheric data to break compositional degeneracies. Recent JWST observations of the temperate sub-Neptune TOI-270~d revealed detections of CH$_4$ and CO$_2$ in its H$_2$-rich atmosphere, with tentative inferences of H$_2$O and CS$_2$ and a non-detection of NH$_3$. We conduct a theoretical exploration of the range of possible interiors for TOI-270~d based on the current observational constraints. We carry out internal structure modelling using a coupled atmosphere-interior model, including self-consistent atmospheric temperature structures informed by JWST observations. The bulk properties permit solutions spanning mini-Neptune, gas dwarf and hycean scenarios, with a wide range of possible surface conditions, which are strongly dependent on the atmospheric properties, including the presence of clouds/hazes. We explore the solutions allowing for surface water oceans on TOI-270~d, including under potentially habitable conditions. The atmospheric mass fractions permitting habitable surface conditions are found to be $\lesssim$$3.5\times10^{-5}$ and pressures $\lesssim$100~bar for the envelope temperature structures considered. We consider mini-Neptune interiors that are sufficiently warm for H$_2$O to be mixed with the H$_2$-rich envelope. Finally, we consider possible gas dwarf interiors, finding H$_2$-rich envelope mass fractions of $\sim$$1-5$\% are required to satisfy the bulk properties, with surface pressures $\sim10^4-10^5$~bar. Further theoretical and experimental studies in addition to future atmospheric observations will aid the characterisation of the possible interior and surface conditions on TOI-270~d.

\end{abstract}

\begin{keywords}
planets and satellites: interiors -- planets and satellites: composition -- planets and satellites: individual: L231-32 -- planets and satellites: exoplanets
\end{keywords}



\section{Introduction} \label{Intro}

A central challenge in exoplanet characterisation is the degeneracy of interior compositions permitted by a planet's bulk properties. This problem is particularly evident within sub-Neptune regime, which encompasses planets with radii $\sim$$1-4\ \mathrm{R_\oplus}$ and is thought to contain a broad range of planets with varying fractions of volatiles. This regime spans rocky super-Earths to volatile-rich sub-Neptunes, with the radius valley separating these two populations \citep[e.g.][]{Fulton2017}. The densities of the larger radius population are consistent with substantial fractions of volatiles and ices, including H$_2$-rich atmospheres and H$_2$O ices. Characterising sub-Neptunes is important for revealing the nature and diversity of planets in this regime which can be related to their formation and evolution, including the shaping of the radius valley, in addition to their potential habitability. Sub-Neptunes orbiting nearby M dwarfs have gained significant attention in the field, as they are both abundant and conducive to atmospheric characterisation with JWST, with observations having been made for multiple such targets \citep[e.g.][]{Madhusudhan2023b,Kempton2023,Cadieux2024,Holmberg2024,Benneke2024,Wallack2024,Madhusudhan2025}. The availability of atmospheric data can begin to rule out possible interior compositions through coupling atmosphere and interior models \citep[e.g.][]{Madhusudhan2020,GuzmanMesa2022,Rigby2024b,Nixon2024}. However, the relationships between the interiors, surfaces and atmospheres of these planets are complex, and remain an active area of study for both water-rich bodies \citep[e.g.][]{Yu2021,Hu2021,Tsai2021,Madhusudhan2023a} and water-poor bodies \citep[e.g.][]{Kite2019,Kite2020,Rigby2024b,Young2024}.

TOI-270~d (L231-32~d) is a temperate sub-Neptune orbiting an M3 host star \citep{Gunther2019,VanEylen2021,Kaye2022}, with planetary mass $M_\mathrm{p}=4.78\pm0.43\ \mathrm{M_\oplus}$ \citep{VanEylen2021} or $4.20\pm0.16\ \mathrm{M_\oplus}$ \citep{Kaye2022} and radius $R_\mathrm{p}=2.133\pm0.058 \ \mathrm{R_\oplus}$ \citep{VanEylen2021}, as given in Table~\ref{table:planetproperties}. Orbiting at $\sim$$0.07$~au \citep{VanEylen2021}, TOI-270~d lies just closer to its host star than the inner edge of the classical habitable zone (HZ), with equilibrium temperature $387$~K ($326$~K) for Bond albedo $A_\mathrm{B}=0$ ($A_\mathrm{B}=0.5$). The bulk density of TOI-270~d is consistent with the presence of an H$_2$-rich atmosphere, evidence of which was reported by \citet{MikalEvans2023}, using transmission spectroscopy with the Hubble Space Telescope Wide Field Camera 3 (HST WFC3). Based on its bulk properties and equilibrium temperature, \citet{Madhusudhan2021} proposed TOI-270~d as a hycean candidate, a class of potentially habitable planet with H$_2$O oceans underlying H$_2$-rich atmospheres, for which the HZ is considerably wider than the classical HZ. 

Recently, TOI-270~d was observed with JWST \citep{Holmberg2024,Benneke2024}. Two studies, \citet{Holmberg2024} and \citet{Benneke2024}, used NIRSpec G395H observations in addition to shorter wavelength coverage with HST and JWST NIRISS respectively, to place constraints on photospheric composition at the terminator via atmospheric retrievals. \citet{Holmberg2024} report strong evidence for CH$_4$ and CO$_2$, with mixing ratios $\sim$$0.1-1\%$, and a non-detection of NH$_3$. This is similar to the reported pattern of detections for the cooler hycean candidate K2-18~b \citep{Madhusudhan2023b}. \citet{Holmberg2024} also report moderate evidence for CS$_2$, and tentative evidence for H$_2$O in the atmosphere of TOI-270~d. They suggest that this planet could be a dark hycean world, due to its relatively high temperature for hycean candidates. The presence of clouds or hazes in the photosphere could not be constrained by the observations, similar to the findings of \citet{MikalEvans2023}. \citet{Benneke2024} also find a lack of NH$_3$ in addition to strong CH$_4$ and CO$_2$ detections, albeit at higher abundances than \citet{Holmberg2024}. They also report a tentative H$_2$O detection, in addition to potential evidence for SO$_2$ and CS$_2$. In contrast to \citet{Holmberg2024} and \citet{MikalEvans2023}, their analysis suggests that the atmosphere is metal-rich, with a mean molecular weight (MMW) of $5.47^{+1.25}_{-1.14}$ amu. The retrieved photospheric temperatures for the terminator differ between \citet{Holmberg2024} and \citet{Benneke2024}. For the \citet{Holmberg2024} dual-transit (DT) case, the $T_0$ is $289^{+80}_{-75}$~K at 10~mbar, while \citet{Benneke2024} find a hotter $T_0$ of $385.3^{+44.2}_{-41.8}$~K for their free-chemistry retrieval at the lower pressure of $0.1$–$1$~mbar.

The NIRSpec and NIRISS data were recently re-analysed by \citet{Felix2025}. In addition to a fiducial model with the same set of chemical species as \citet{Benneke2024}, they consider models including an expanded set of molecules, including CH$_3$F, CH$_3$Cl, and/or sulfur chemistry with H$_2$CS, CS, and (CH$_3$)$_2$S (dimethyl sulfide, DMS), all of which were found to be preferred over their fiducial model, but could not be distinguished. They also report detections of CH$_4$ and CO$_2$, while the inclusion of H$_2$O is not preferred by the retrieval. Similarly to \citet{Benneke2024}, they find a high mean molecular weight atmosphere. The chemical species with retrieved abundances most significantly different between all three studies is CH$_4$, with \citet{Benneke2024} finding a high mixing ratio of $\log(X_{\mathrm{CH_4}})=-1.64_{-0.36}^{+0.38}$. \citet{Holmberg2024} find a significantly lower abundance, at $\log(X_{\mathrm{CH_4}})=-2.72_{-0.50}^{+0.41}$ for the one offset + dual transit case. Two of the seven retrieval cases of \citet{Felix2025} give a CH$_4$ abundance consistent with \citet{Benneke2024} to within $1\sigma$, while six of the seven cases have CH$_4$ abundance consistent with \citet{Holmberg2024} -- the exception is the fiducial model, which is disfavoured. 

Recently, \citet{Glein2025} conducted thermochemical equilibrium calculations to investigate the quenching of C-H-O-N species in the atmosphere of TOI-270~d. This study adopted the atmospheric abundances of \citet{Benneke2024}. The methods outlined in this work for the CO$_2$-CH$_4$ speciation are reported to be applicable only for a lack of water clouds. They suggest that a lack of detected CO can be explained by equilibrium chemistry of hot gas, negating the need to invoke a hycean scenario for TOI-270~d and K2-18~b. This would be potentially the case for an H$_2$O-rich and CO$_2$-poor atmosphere. This argument hinges on the high mean molecular weight (MMW) found by \citet{Benneke2024}, and subsequently \citet{Felix2025}. 

The interpretation of precise chemical abundances in sub-Neptune atmospheres from JWST requires further studies into the complex relationship between sub-Neptune atmospheres, surfaces and interiors. This is evident from the case of habitable zone sub-Neptune K2-18~b, which was recently observed with JWST \citep{Madhusudhan2023b}. The abundance of CH$_4$ and CO$_2$ in addition to the non-detection of NH$_3$ was suggested to indicate the presence of a surface ocean (i.e. hycean conditions) on K2-18~b \citep{Madhusudhan2023b}, based on chemical models \citep{Hu2021,Madhusudhan2023a}. The other interior solutions permitted by the bulk properties were shown to span a rocky world with thick H$_2$-rich envelope and a mini-Neptune scenario \citep{Madhusudhan2020}, as is the case for similar density sub-Neptunes like TOI-270~d.

Additional photochemical modelling by \citet{Cooke2024} found that a mini-Neptune scenario is incompatible with the retrieved atmospheric abundances for K2-18~b, in contrast to previous studies \citep{Wogan2024}. An uninhabited hycean scenario was found to explain the abundance constraints with the exception of CH$_4$ \citep{Cooke2024}. The equivalent inhabited hycean case was found to better match observations, with CH$_4$ produced by methanogenic life satisfying the high observed CH$_4$ abundance. Recent work explored the possibility of a gas dwarf scenario for K2-18~b. \citet{Rigby2024b} reported a framework including internal structure modelling, atmospheric structure modelling, melt-atmosphere interactions, photochemical modelling and spectral predictions. They find that the CO/CO$_2$ ratio for K2-18~b and the lack of nitrogen are incompatible with predictions for a gas dwarf scenario. However, the relationship between planetary surfaces and observable atmospheres requires the use of models that rely on incomplete data \citep[e.g.][]{Shorttle2024,Rigby2024b,Glein2025} -- for instance, for the solubility of chemical species in a magma ocean at the relevant high pressures and temperatures, and the uncertain phase behaviour of planetary materials. Atmospheric data remains a key factor in breaking degeneracies between possible interior compositions. However, we have not yet reached a position where robust constraints can be made on the nature of a sub-Neptune's surface via atmospheric observations. 

\setlength{\arrayrulewidth}{1.1pt}
\begin{table*}
\begin{tabular}{  wc{2.4cm}   wc{1.4cm}   wc{1.5cm}   wc{1.2cm}   wc{1.2cm}   wc{1.2cm} wc{1.2cm}  }
     \hline
     \textbf{Source} & $\mathbf{M_\mathbf{p}/\mathrm{\mathbf{M}}_{\oplus}}$ & $\mathbf{R_{p}/\mathrm{\mathbf{R}}_{\oplus}}$ & \textbf{$\mathbf{T_{eq,0}}$ /K} & \textbf{$\mathbf{T_{eq,0.5}}$ /K} & \textbf{$\mathbf{a}$/AU} & \textbf{$\mathbf{Period}$/$\mathbf{days}$} \\
     \hline

     \citet{Gunther2019} & & 2.13$\pm0.12$ & 372 & 313 & 0.073 & 11.4 \\

     \citet{VanEylen2021} & 4.78$\pm0.43$ & 2.133$\pm0.058$ & 387 & 326 & 0.072 & 11.4 \\ 

     \citet{Kaye2022} & 4.20$\pm0.16$ &  &  &  &  &   \\   

     \citet{MikalEvans2023} &  &  $2.19\pm0.07$ &   &   &   &    \\ 

     \hline
\end{tabular}
\caption{Reported properties of TOI-270~d. Equilibrium temperature values are calculated with $A_\mathrm{B}=0$ and $A_\mathrm{B}=0.5$, and assuming uniform day-night redistribution.}
\label{table:planetproperties}
\end{table*}

In this study we present an exploration of the range of possible interior compositions and surface conditions for the temperate sub-Neptune TOI-270~d based on the available atmospheric constraints from JWST observations \citep{Holmberg2024}. We use atmospheric pressure-temperature profiles generated through self-consistent atmospheric modelling, based on retrieved atmospheric abundances \citep{Holmberg2024} to place constraints on the possible interior compositions that satisfy the bulk properties of the planet. These span a wide range of possible water mass fractions and surface conditions, including gas dwarfs, mini-Neptunes and hycean worlds. We explore the solutions that permit liquid water at the planet surface, including habitable hycean conditions, and place constraints on the possible ocean depths. We perform initial modelling of dark hycean scenarios for TOI-270~d, where inefficient day-night redistribution could lead to a habitable nightside for a dayside that is too hot for life. We investigate mini-Neptune scenarios, including mixed H$_2$O/H$_2$ both with and without a cold trap which would result in an H$_2$-rich layer above the mixed portion of the envelope. We explore gas dwarf scenarios, placing constraints on the permitted H$_2$-rich envelope mass fractions and surface conditions at the envelope/rock interface. Finally, we discuss the future prospects for the characterisation of TOI-270~d via atmospheric observations coupled to atmosphere and interior models, and the avenues for future research required to robustly identify the nature of this and similar planets.

\section{Methods} \label{Methods}

We first outline our procedure for calculating the possible interior compositions and surface conditions of TOI-270~d. We use the planetary bulk properties and self-consistently generated atmospheric pressure-temperature ($P$-$T$) profiles informed by JWST observations \citep{Holmberg2024} to place constraints on the interior composition of the planet. In this section we describe the internal structure model HyRIS, outlined in full in \citet{Rigby2024}. We detail the assumptions made, including the equations of state and temperature profile used for each planetary component layer. We refer the reader to \citet{Rigby2024} and \citet{Rigby2024b} for further description of the internal structure model and equations of state.

\subsection{Internal Structure Model} \label{Method:IS}

A range of studies have developed internal structure models to relate exoplanet observables to their interior compositions \citep[e.g.][]{Leger2004,Fortney2007,Seager2007,Sotin2007,Valencia2007,Rogers2010a,Madhusudhan2012,Zeng2013,Thomas2016a,Dorn2017,Madhusudhan2020,Nixon2021,Huang2022,Rigby2024}. We use the internal structure model HyRIS, outlined in \citet{Rigby2024}, to relate the planetary mass and radius to possible interior compositions. The model is applicable to sub-Neptunes and is specialised for water-rich interiors, including surface oceans. \citet{Rigby2024} used HyRIS configured with four differentiated planetary layers: H$_2$-rich envelope, H$_2$O layer, silicate (MgSiO$_3$ perovskite) mantle and iron (Fe) core, similar to previous studies of sub-Neptune interiors \citep[e.g.][]{Rogers2010a,Nixon2021}. In this work, we also consider mixed envelopes containing H$_2$O in addition to H$_2$ and He, and a more complex treatment of silicates, as in \citet{Rigby2024b}. 

The internal structure model outputs the planet radius $R_\mathrm{p}$ from inputs of planet mass $M_{\mathrm{p}}$, the mass fractions of the constituent layers $x_{\mathrm{i}} = M_{\mathrm{i}}/M_{\mathrm{p}}$, and the photospheric pressure and temperature, $P_0$ and $T_0$. Under the assumption of spherical symmetry, the model solves the equations for mass continuity, 

\begin{equation}
    \frac{dR}{dM} = \frac{1}{4\pi R^{2}\rho(P,T)}
    \label{eqn:masscontinuity}
\end{equation}

\noindent
and hydrostatic equilibrium,

\begin{equation}
    \frac{dP}{dM} = -\frac{GM}{4\pi R^{4}}
    \label{eqn:hydrostaticeq}
\end{equation}

\noindent
where $R$ is the radius of the spherical shell, $M$ is the mass enclosed, $\rho$ is the density and $P$ is the pressure. These equations are solved via fourth-order Runge-Kutta numerical integration with variable step size, with boundary conditions at the planet's exterior surface. A bisection root-finding procedure is used to obtain $R_{\mathrm{p}}$. The convergence condition is defined to be a central radius with zero enclosed mass of $0 < R(M=0) < 100\ \mathrm{m}$. See \citet{Rigby2024} for a more detailed description of the model, and their Figure 1 for a diagram of the model architecture. For each planetary layer, an equation-of-state (EOS) is required, along with a description of the $P$-$T$ structure. The EOS describes the density as a function of pressure and temperature, $\rho = \rho(P,T)$. The choice of temperature profiles, $T=T(P)$, and EOS for each layer are outlined in Sections \ref{Method:PTprofiles} and \ref{Method:EOSs}. 

In Section \ref{Res:Interiorcomps}, the internal structure model is evaluated for a grid of models across the full phase space of mass fraction combinations, for different assumptions of core composition and atmospheric $P$-$T$ profile. (We use ``core'' to refer to the silicate and iron layers collectively, such that $x_{\mathrm{core}}=x_\mathrm{silicate}+x_\mathrm{Fe}$.) The inputted $M_\mathrm{p}$ is also varied within its $1\sigma$ uncertainty. We then extract the interior solutions that satisfy the measured $R_\mathrm{p}$ to within the $1\sigma$ uncertainty, defined via an ellipse centred on $M_\mathrm{p}$ and $R_\mathrm{p}$. In addition to $R_\mathrm{p}$, the model outputs the internal structure and phase structure. This facilitates analysis of the surface conditions and values including the ocean or magma ocean depth where relevant, as described in \citet{Rigby2024} and \citet{Rigby2024b}. 

\subsection{Temperature Profiles} \label{Method:PTprofiles}

Under the assumption of vigorous convection, we assume an adiabatic temperature profile in the interior below the envelope. The adiabatic profiles are described by the adiabatic temperature gradient,

\begin{equation}
    \left. \frac{\partial T}{\partial P}\right|_{S} = \frac{\alpha T}{\rho c_p}
    \label{eqn:adiabat}
\end{equation}

\noindent
where $\alpha$ is the coefficient of volume expansion and $c_p$ is the isobaric specific heat capacity. $\alpha$ is derived from the EOS,

\begin{equation}
    \alpha = \frac{1}{V} \left. \frac{\partial V}{\partial T}\right|_{P} = - \left. \frac{\partial \ln{\rho}}{\partial T}\right|_{P}
    \label{eqn:alpha}
\end{equation}

\noindent
where $V$ is the specific volume.  

\begin{figure}
    \centering
    \includegraphics[width=0.8\columnwidth]{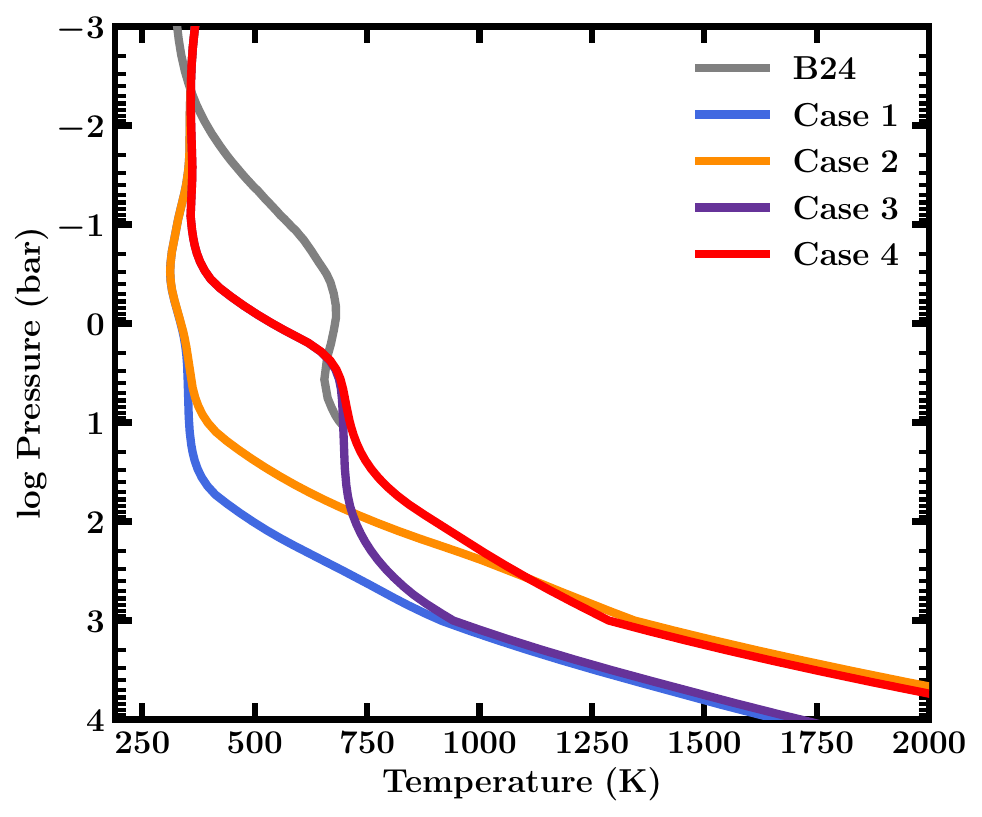}
    \caption{Temperature structures adopted for the envelope, generated with GENESIS using the retrieved abundances of \citet{Holmberg2024}. These cases correspond to each combination of $T_\mathrm{int}$ of $25$ and $50$~K, and Rayleigh enhancement factor $a$ of $100$ and $1500$. The profiles, generated to $1000$~bar, are extended using an adiabat. We also show the evening $P$-$T$ profile from Figure 10 of \citet{Benneke2024}, which is only provided to 10~bar pressures.}
    \label{fig:PTenvelope}
\end{figure}

For the H$_2$-rich envelope, we use pressure-temperature ($P$-$T$) profiles generated via self-consistent atmospheric modelling. This modelling is informed by the retrieved atmospheric parameters for TOI-270~d from JWST data \citep{Holmberg2024}. We use the GENESIS framework \citep{Gandhi2017,Madhusudhan2020,Piette2020,Madhusudhan2021,Madhusudhan2023a}, which assumes a plane parallel atmosphere and conducts line-by-line radiative transfer via the Feautrier method, and solves for radiative-convective equilibrium via the Rybicki linearization scheme -- see the references above for a more detailed description of the method. For the chemical composition we assume values according to the median retrieved values from \citet{Holmberg2024}; we adopt $\log{X_{\mathrm{H_2O}}}=-2.0$, $\log{X_{\mathrm{CH}_4}}=-2.5$ and $\log{X_{\mathrm{CO}_2}}=-2.5$. The other key inputs are the internal temperature $T_\mathrm{int}$, the properties of the host star, the incident irradiation, day-night energy redistribution, and cloud/haze properties. We consider values of $25$~K and $50$~K for $T_\mathrm{int}$, similar to previous studies \citep{Madhusudhan2020}, and values for the Rayleigh enhancement factor $a$ for hazes of $100$ and $1500$. The resulting $P$-$T$ profiles are generated to $1000$~bar pressures. The Rayleigh enhancement factor allows a parameterisation of hazes as H$_2$ Rayleigh scattering multiplied by some factor $a$. These values of $a$ were chosen to explore a wide range of atmospheric temperature structures which could be permitted by current observations, and are similar to those used in previous studies \citep[e.g.][]{Piette2020,Madhusudhan2021,Rigby2024b}. We extrapolate the $P$-$T$ profiles to higher pressures using an adiabat, calculated using the parameters from \citet{Chabrier2019} for solar proportion H/He. The resulting profiles are shown in Figure \ref{fig:PTenvelope}. Case 1, the coldest profile, has $a=1500$ and $T_\mathrm{int}=25$~K; Case 2 has $a=1500$ and $T_\mathrm{int}=50$~K; Case 3 has $a=100$ and $T_\mathrm{int}=25$~K; and Case 4 has $a=100$ and $T_\mathrm{int}=50$~K. For all cases, excluding the mini-Neptune scenarios considered in Section~\ref{Res:MiniNep}, the density of the envelope is calculated assuming $10\%$ H$_2$O by mass, informed by the observed H$_2$O mixing ratio. A full description of the EOS for this is given below. 

\subsection{Equations of state} \label{Method:EOSs}

We explore internal structures with different combinations of an iron core, a silicate mantle, a pure H$_2$O layer, and an H$_2$-rich envelope. We also consider the possibility of a mixed H$_2$/H$_2$O envelope. The equations of state for the different components are as follows.

\subsubsection{H$_2$-rich and mixed H$_2$/H$_2$O envelopes} \label{Methods:HEOS}

We use the EOS of \citet{Chabrier2019} to describe the H$_2$-rich envelope. This EOS is for H/He with a solar helium fraction ($Y=0.275$) -- see \citet{Chabrier2019} for a more detailed description of its derivation. In this work, as described above, we consider the atmospheric abundances of \citet{Holmberg2024} and hence include $10\%$ H$_2$O by mass in the envelope. Our EOS for H$_2$O is described in Section~\ref{Methods:H2OEOS}. 

The EOS for a mixed envelope is calculated by combining the densities proportionally to the respective mass fractions ($z_\mathrm{i}$) for each component,

\begin{equation}
    \rho_\mathrm{mix}(P,T) = \left[{\sum \frac{z_i}{\rho_i(P,T)}}\right]^{-1}
    \label{eqn:mixeddensity}
\end{equation}

and the adiabatic gradient is calculated via 

\begin{equation}
    \left( \left. \frac{\partial \mathrm{log} T}{\partial \mathrm{log} P} \right|_S \right)_\mathrm{mix} = - \frac{\sum_i z_i S_i \left. \frac{\partial \mathrm{log} S_i}{\partial \mathrm{log} P} \right|_T
}{\sum_i z_i S_i \left. \frac{\partial \mathrm{log} S_i}{\partial \mathrm{log} T} \right|_P
}
    \label{eqn:mixedadgrad}
\end{equation}

where $S$ is the entropy. As for the density, we use the entropy and entropy derivatives with respect to pressure and temperature from \citet{Chabrier2019}. The sources for the H$_2$O components are given below. 

\subsubsection{H$_2$O} \label{Methods:H2OEOS}

Studies including H$_2$O-rich interiors require a careful treatment of the complex phase structure of H$_2$O \citep[e.g.][]{Thomas2016a,Mousis2020,Huang2021,Haldemann2020,Nixon2021,Rigby2024}. We use a temperature-dependent EOS for pure H$_2$O, outlined in full in \citet{Rigby2024}, which was compiled following a similar approach to \citet{Thomas2016a} and \citet{Nixon2021}. The EOS is valid for temperatures in the range $200 - 24000$ K and pressures $10^2 - 10^{22}$ Pa ($10^{-3}-10^{17}$ bar). It is comprised of a number of different sources valid for certain phases and/or regions of $P$-$T$ space. We outline these briefly here, and are described in more detail in \citet{Rigby2024}. For the vapour, liquid and most of the supercritical regions we use the IAPWS-1995 formulation from \citet{Wagner2002}. Ice Ih is covered by data from \citet{Feistel2006}. The results of \citet{French2009} are used for regions spanning supercritical, ice VII, ice X and superionic ice. The majority of the ice VII phase is described by a Vinet EOS with a thermal correction \citep{Fei1993}, and ice VIII is covered by an extrapolated form of this EOS \citep{Klotz2017}. Ices II, III, V and VI are described by the relevant EOS from \citet{Journaux2020b}. A Thomas-Fermi-Dirac (TFD) EOS is used for high-pressure regions \citep{Salpeter1967}. Remaining high-pressure regions spanning the ice VII to X transition are covered by the H$_2$O EOS of \citet{Seager2007}. For any remaining regions, which is largely supercritical, we extrapolate the IAPWS-1995 formulation \citep{Wagner2002}. 

The phase-boundaries we use are from \citet{Dunaeva2010}, with the exception of the liquid-vapour boundary from \citet{Wagner2002}. The full H$_2$O EOS and phase-diagram are shown in Figure 2 in \citet{Rigby2024}. For the isobaric specific heat capacity $c_\mathrm{p}$, required for the adiabatic gradient (Equation~\ref{eqn:adiabat}), we use the same sources as the EOS where available \citep{Wagner2002,Feistel2006,Journaux2020a}. For regions where there is no data for $c_\mathrm{p}$, we adopt the value of the nearest available point in $P$-$T$ space. The coefficient of volume expansion, $\alpha$, also required for calculations of the adiabat, is calculated directly from the EOS. 

For H$_2$O miscible in H$_2$, we require the entropy and its derivatives with respect to pressure and temperature to calculate the adiabatic gradient for the mixture (Equation~\ref{eqn:mixedadgrad}). The entropy (only required for the vapour and supercritical phases) is calculated using the IAPWS-1995 formulation \citep{Wagner2002}. 

\subsubsection{Silicate mantle and iron core}

In order to explore gas dwarf scenarios for TOI-270~d, we consider a treatment of the silicate mantle following \citet{Rigby2024b}. We consider temperature-dependent EOSs based on experimental data for molten and solid silicates, adopting a peridotitic composition. In this work, as in \citet{Rigby2024b}, we assume full melting occurs at the peridotite liquidus, and hence, for simplicity, do not include a prescription for partial melting of peridotite. The liquidus is adopted from \citet{Fiquet2010} and \citet{Monteux2016}. 

For molten peridotite, we follow the approach of \citet{Monteux2016}. We combine by mass fraction the densities of molten enstatite (MgSiO$_3$, $33\%$), forsterite (Mg$_2$SiO$_4$, $56\%$), fayalite (Fe$_2$SiO$_4$, $7\%$), anorthite (CaAl$_2$Si$_2$O$_8$, $3\%$) and diopside (MgCaSi$_2$O$_6$, $1\%$). The densities of the individual molten minerals are described by third-order Birch-Murnaghan/Mie-Gruneisen (BM/MG) EOSs from \citet{Thomas2013}. The adiabatic gradient is calculated using Equation \ref{eqn:adiabat}, where $\alpha$ is calculated directly from the EOS using Equation \ref{eqn:alpha} and $c_\mathrm{p}$ is adopted from \citet{Monteux2016}. 

For solid peridotite, we use the EOS of \citet{Lee2004}, which also takes the form of a BM/MG EOS for the constituent minerals (orthorhombic perovskite ($64\%$), magnesiow\"ustite ($31\%$) and calcium perovskite ($5\%$)), and are similarly weighted by mass. For the adiabatic gradient, $\alpha$ is calculated from the EOS and we adopt the $c_\mathrm{p}$ value from \citet{Lee2004}. The limit of the experimental data for this EOS is $1.07\times10^{11}$~Pa. Beyond these pressures, we use the EOS of \citet{Seager2007} for MgSiO$_3$ perovskite. This EOS is temperature-independent and was originally derived at $300$~K. We note that at the relevant pressures, the thermal effects in solid silicates are expected to have negligible consequence for the internal structure \citep{Seager2007,Grasset2009}. The \citet{Seager2007} EOS is a fourth-order Birch-Murnaghan EOS \citep{Birch1952,Karki2000}, and at high pressures a TFD EOS \citep{Salpeter1967}. 

The silicate mantle below a high-pressure ice layer would typically exist at pressures $>1.07\times10^{11}$~Pa, beyond pressures covered by the \citet{Lee2004} EOS. In water-rich cases (i.e. containing sufficient water to be present as ices), we hence use the \citet{Seager2007} EOS only, as in \citet{Rigby2024}.

For the iron core, we adopt the EOS from \citet{Seager2007} for hexagonal close-packed Fe. This EOS is also temperature-independent, and is formed of a Vinet EOS \citep{Vinet1989,Anderson2001} and a TFD EOS at high pressures.

\section{Results} \label{Results}

In this section, we present our results for the possible interior compositions of TOI-270~d based on the planetary bulk properties and constraints from recent transmission spectroscopy with JWST \citep{Holmberg2024}. As described in Section \ref{Method:PTprofiles}, we adopt self-consistent temperature profiles in the envelope, generated using the atmospheric abundances from \citet{Holmberg2024}. We investigate the range of interior conditions possible for TOI-270~d with solutions spanning mini-Neptunes, rocky gas dwarfs with thick H$_2$-rich envelopes, and hycean worlds. In Section~\ref{Res:Obs}, we first describe the observational constraints. In the subsequent sections, we then explore each class of possible interior for TOI-270~d. 

\subsection{Observational constraints} \label{Res:Obs}

\begin{figure}
    \centering
    \includegraphics[width=\columnwidth]{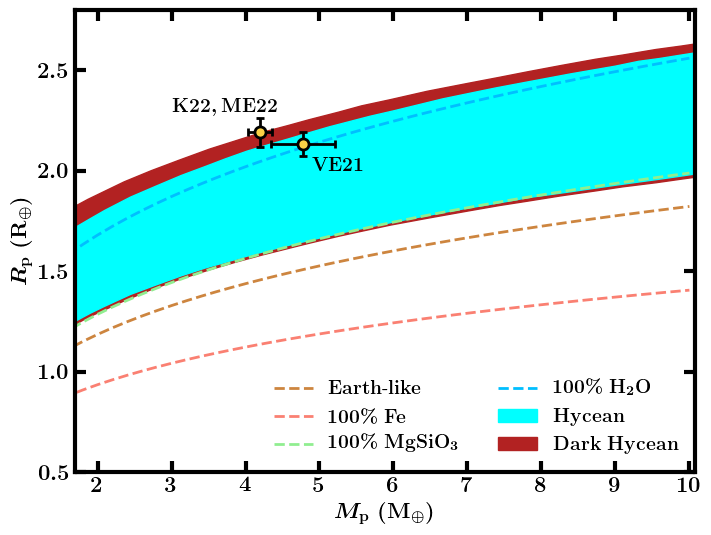}
    \caption{Mass-radius curves for different planet compositions, including the standard isothermal $M$-$R$ curves shown by dashed lines. The hycean mass-radius plane from \citet{Madhusudhan2021} is also shown. The mass and radius measurements for TOI-270~d are shown. VE21 is \citet{VanEylen2021}, ME22 is \citet{MikalEvans2023} and K22 is \citet{Kaye2022}.}
    \label{fig:MR}
\end{figure}

In Figure \ref{fig:MR} we show the reported mass ($M_\mathrm{p}$) and radius ($R_\mathrm{p}$) measurements for TOI-270~d alongside mass-radius ($M$-$R$) relations \citep{Seager2007}, which include homogeneous compositions of H$_{2}$O, silicates and iron, in addition to an Earth-like rocky composition (all isothermal at $300$~K). The hycean $M$-$R$ plane from \citet{Madhusudhan2021} is also shown. TOI-270~d lies towards the lower density region of the hycean $M$-$R$ plane and within the dark hycean region, depending on the $M_\mathrm{p}$ and $R_\mathrm{p}$ values adopted. In this study we consider those of \citet{VanEylen2021} for consistency with \citet{Holmberg2024}.

Using the JWST NIRSpec G395H transmission spectrum in addition to HST data \citep{MikalEvans2023}, \citet{Holmberg2024} report strong evidence for CH$_4$ and CO$_2$ and a non-detection of NH$_3$, in addition to tentative evidence for H$_2$O and CS$_2$. For the \citet{Holmberg2024} dual-transit (DT) case, the retrieved $T_0$ is $289^{+80}_{-75}$~K at 10~mbar. This study, along with \citet{MikalEvans2023}, reports a low MMW H$_2$-dominated atmosphere. In contrast, \citet{Benneke2024} and \citet{Felix2025} report a high MMW atmosphere, but with \citet{Felix2025} finding a CH$_4$ abundance broadly consistent with \citet{Holmberg2024}. In this study, we consider the atmospheric abundances of the published work, \citet{Holmberg2024}. The additional JWST observations of TOI-270~d may provide the tools to confirm the cause of the discrepancy between these studies, and place more robust constraints on the photospheric abundances for TOI-270~d. We note that the set-up of atmospheric retrievals remains crucial to accurately constrain the atmospheric abundances and photospheric temperature. 

In Sections \ref{Res:Interiorcomps}-\ref{Res:Rocky} we consider the $P$-$T$ profiles Cases 1-4, based on the findings of \citet{Holmberg2024}, to explore the range of possible interior scenarios for TOI-270~d. In Figure \ref{fig:PTenvelope} we show these $P$-$T$ profiles, generated self-consistently with GENESIS, as described in Section \ref{Method:PTprofiles}, using the retrieved abundances of \citet{Holmberg2024}. Cases 1 through 4 correspond to values of $T_\mathrm{int}=25$~K, $a=1500$; $T_\mathrm{int}=50$~K, $a=100$; $T_\mathrm{int}=25$~K, $a=100$; $T_\mathrm{int}=50$~K, $a=100$ respectively, as described in Section \ref{Method:PTprofiles}. We also show a $P$-$T$ profile from \citet{Benneke2024}, generated via a dual-grey GCM. Specifically, we show the ``evening'' profile from their Figure 10, however the profiles in this figure are very similar and do not differ much at pressures above $\sim$1~mbar. This profile is only given to $10$~bar. 

The envelope $P$-$T$ profiles Cases 1-4 can be used to estimate the potential surface conditions possible, which is particularly informative for H$_2$O-rich compositions. Our $P$-$T$ profiles are shown against the phase diagram for H$_2$O in Figure~\ref{fig:PTh2o}. Cases 1 \& 2 cross the vaporisation curve and therefore have the potential to permit liquid water at the surface, if the corresponding composition producing these surface conditions is compatible with the bulk density. We show that such solutions are possible in Section~\ref{Res:Hycean}. 

On the other hand, Cases 3 \& 4 lack the presence of a cold trap, and do not cross the vaporisation curve. This implies that the H$_2$/H$_2$O boundary (HHB) must fall either in the vapour or supercritical region of the H$_2$O phase diagram. According to studies investigating the miscibility of H$_2$O and H$_2$, \citep[e.g.][]{Gupta2024}, this could result in a mixed H$_2$/H$_2$O envelope \citep{Nixon2021,Benneke2024}. In this scenario of a fully miscible envelope the H$_2$ and H$_2$O form a homogenous mixture, with each component present in any proportion. For reference, in Figure~\ref{fig:PTh2o} we show the critical curve from \citet{Gupta2024}, representing the transition from two separate phases of H$_2$ and H$_2$O to one phase. A mixed envelope would not be compatible with the median observed mixing ratios for H$_2$O, including for the canonical DT case \citep{Holmberg2024}. We note that with further observation, a higher observed H$_2$O mixing ratio, for instance towards the upper limit of the error bar on current estimates, may permit a fully mixed H$_2$O/H$_2$ envelope.

\subsection{Degeneracy in interior composition} \label{Res:Interiorcomps}

\begin{figure}
    \centering
    \includegraphics[width=\columnwidth]{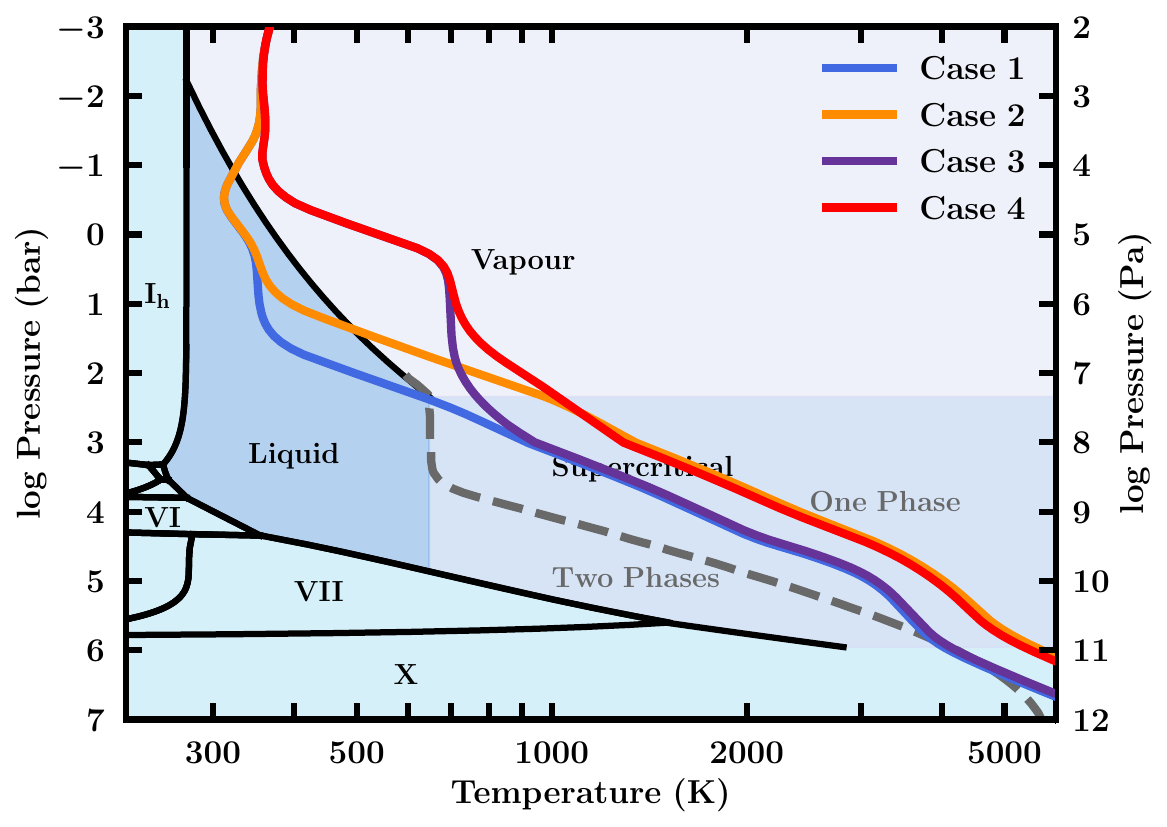}
    \caption{Model atmospheric $P$-$T$ profiles relative to the phase diagram for a 100\% H$_2$O surface below the atmosphere. At a given point on a $P$-$T$ profile the phase of H$_2$O in the background corresponds to the phase of a pure H$_2$O surface assuming that point represents the boundary between the atmosphere and a pure H$_2$O interior \citep[following][]{Madhusudhan2020, Piette2020, Rigby2024}. Thus, each $P$-$T$ profile effectively represents the locus of pressure and temperature values permissible at the boundary between the atmosphere and the interior for that model. The H$_2$O phase boundaries for pure H$_2$O are obtained from \citet{Dunaeva2010}. The critical curve for H$_2$/H$_2$O mixtures is shown in the grey dashed line, above which H$_2$ and H$_2$O would be expected to exist as one phase \citep{Gupta2024}. The atmospheric $P$-$T$ profiles are generated using self-consistent models as discussed in Section~\ref{Method:PTprofiles}.}
    \label{fig:PTh2o}
\end{figure}

\begin{figure*}
    \centering
    \includegraphics[width=0.7\columnwidth]{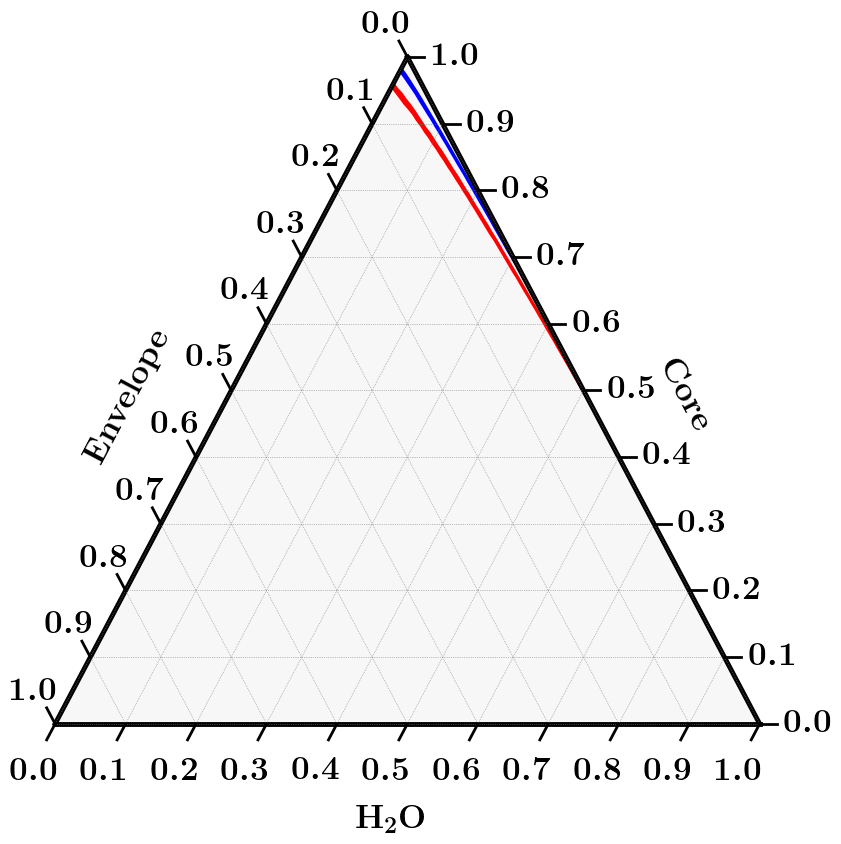}
    \includegraphics[width=0.99\columnwidth]{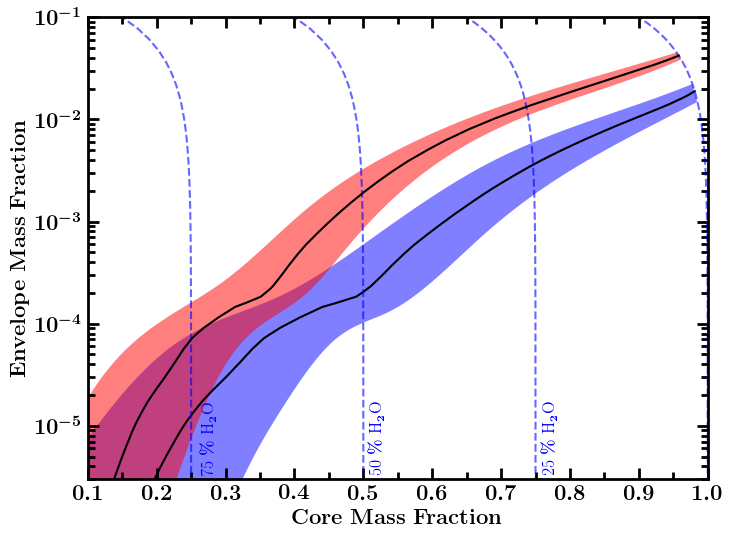}
    \caption{Left: Ternary diagram showing the best-fit compositions for TOI-270~d, for an Earth-like core composition (blue) and a $100\%$ Fe core (red). In both cases we adopt Case 1 as the atmospheric $P$-$T$ profile. Right: Mass fractions of the H$_2$-rich envelope against core mass fraction, for Earth-like (blue) and $100\%$ Fe (red) core compositions. The red and blue shaded regions indicate combinations that reproduce the observed mass and radius to within $1\sigma$. The best fit solutions in each case are shown by the black solid line. The dashed blue lines show constant H$_2$O mass fractions at $25\%$ intervals.} 
    \label{fig:ternarycontour}
\end{figure*}

We firstly place constraints on the range of possible interior compositions that can reproduce the $M_{\mathrm{p}}$ and $R_{\mathrm{p}}$ of TOI-270~d to within their $1\sigma$ uncertainties. For consistency with \citet{Holmberg2024}, we adopt $M_{\mathrm{p}}=4.78\pm0.43\ \mathrm{M_{\oplus}}$ and $R_{\mathrm{p}}=2.133\pm0.058\ \mathrm{R_{\oplus}}$ \citep{VanEylen2021}. Initially, two end-member core compositions are considered -- the extreme case of a high density $100\%$ Fe core ($x_\mathrm{core}=x_\mathrm{Fe}$, $f_\mathrm{silicate}=0$), and an Earth-like composition core ($f_\mathrm{silicate}=67\%$), where we have defined $f_\mathrm{silicate}$ to be the silicates-to-iron ratio in the interior. We vary the envelope, H$_2$O and core mass fractions across the full parameter space, taking an extreme value for the maximum $x_\mathrm{H_2O}$ of $95\%$. In this and following sections, we consider the $P$-$T$ profiles informed by the \citet{Holmberg2024} results -- these are Cases 1-4, generated via self-consistent modelling, as described in Section \ref{Method:PTprofiles}. Case 1 is adopted as our canonical profile, as it permits the full range of possible interior scenarios.

The interior solutions able to reproduce the measured $M_\mathrm{p}$ and $R_\mathrm{p}$ of TOI-270~d to within $1\sigma$ are shown in Figure~\ref{fig:ternarycontour}, for core compositions of pure Fe and Earth-like, and $P$-$T$ profile Case 1. In the left-hand plot we represent these solutions on a ternary diagram, with the coloured regions encompassing the $1\sigma$ solutions for each of the considered core compositions. Red corresponds to pure Fe while blue corresponds to an Earth-like core composition. In the right-hand side of Figure \ref{fig:ternarycontour} we show the envelope mass fraction against the core mass fraction for the $1\sigma$ solutions. The shaded regions show the $1\sigma$ solutions as in the ternary diagram, while the black line indicates the best-fit solutions that reproduce the median $M_{\mathrm{p}}$ and $R_{\mathrm{p}}$. The blue dashed lines indicate a constant H$_2$O mass fraction. In these initial calculations, we have assumed that the H$_2$O layer is distinct, i.e. is not miscible in the H$_2$-rich envelope, beyond the standard $10\%$ by mass as informed by observations.

As evident from these figures, the data permit a wide range of envelope, H$_2$O and hence core mass fractions. These include interior solutions with H$_2$O mass fractions ($x_\mathrm{H_2O}$) across the full range considered ($0-95\%$). Within the $1\sigma$ solutions, a larger $x_\mathrm{H_2O}$ generally corresponds to a smaller envelope mass fraction ($x_\mathrm{env}$), as is evident in Figure~\ref{fig:ternarycontour}. The upper limit of $x_{\mathrm{env}}$ is constrained to be $4.62\%$, for an unrealistic high-density $100\%$ Fe core and no distinct H$_2$O layer. For lower density core compositions, the maximum envelope mass fraction is lower. For an Earth-like core composition, the maximum envelope fraction is found to be $2.27\%$.

The choice of envelope $P$-$T$ profile affects the permitted interior solutions and surface conditions. For instance, as is evident from Figure \ref{fig:PTh2o}, only Cases 1 and 2 would permit a liquid ocean. Additionally, as described above, Cases 3 and 4 may not be compatible with the observed atmospheric abundances due to the miscibility of the H$_2$O in the H$_2$-rich envelope for a HHB lying in the vapour or supercritical region of the H$_2$O phase diagram. For interiors lacking a distinct H$_2$O layer, i.e. gas dwarf solutions, the choice of $P$-$T$ profile affects the envelope mass fraction required. For instance, with an Earth-like interior, the maximum $x_\mathrm{env}$ for Case 1 is $2.27\%$. For Case 2, which is significantly hotter in the deep atmosphere due to the higher $T_\mathrm{int}$, the maximum $x_\mathrm{env}$ reduces to $1.33\%$ for an Earth-like interior. The overall hottest profile, Case 4, has the lowest maximum $x_\mathrm{env}$ of $1.06\%$ for an Earth-like composition. Case 3, with the lower $T_\mathrm{int}=25$~K and low $a$, has intermediate maximum $x_\mathrm{env}=1.61\%$ for an Earth-like core. We note that none of our atmospheric $P$-$T$ profiles permit an immisicible supercritical H$_2$O ocean, as explored by \citet{Luu2024} for K2-18~b. As seen in Figure~\ref{fig:PTh2o}, our $P$-$T$ profiles lie above the critical curve from \citet{Gupta2024}, indicating that the H$_2$O and H$_2$ would likely be miscible at these pressure/temperature HHBs, hence precluding an immiscible supercritical H$_2$O ocean. Cases with mixed H$_2$O/H$_2$ are modelled in Section \ref{Res:MiniNep}.

In Figure \ref{fig:HHBlocus} we show the locus of $P_{\mathrm{HHB}}$ and $T_{\mathrm{HHB}}$ for the possible interior solutions of TOI-270~d, including cases with an Earth-like and a pure Fe core, assuming the Case 1 $P$-$T$ profile in the envelope, corresponding to the solutions in Figure~\ref{fig:ternarycontour}. The locus traces the envelope $P$-$T$ profile between lower and upper limits of $T_\mathrm{HHB}=$ 312~K and $P_\mathrm{HHB}=$ 0.3~bar, and $T_\mathrm{HHB}=$ 2922~K and $P_\mathrm{HHB}=7.4\times10^{4}$~bar. The lower pressure and temperature HHBs generally correspond to smaller envelope mass fractions, and hence larger H$_2$O mass fractions, as in Figure \ref{fig:ternarycontour}. 

A wide range of compositions for TOI-270~d are permitted, for the $P$-$T$ profiles Cases 1-4, generated using the abundances of \citet{Holmberg2024}. Based on these abundances, these interiors generally have hydrogen-dominated envelopes. Broadly, the interior solutions fall into three categories -- hycean worlds (with liquid water oceans), mini-Neptunes, and gas dwarfs -- as shown in Figure~\ref{fig:cases}. These classes of solution will be explored in detail in subsequent Sections \ref{Res:HyceanOverall}-\ref{Res:Rocky}. We also show a mixed envelope case, representing the suggested interior of \citet{Benneke2024}. In the following, where a distinct boundary occurs between the H$_2$O and H$_2$, we refer to the boundary as the H$_2$O/H$_2$-envelope boundary, or HHB \citep{Madhusudhan2020}. The equivalent for gas dwarf cases we refer to simply as the surface, or atmosphere-rock/magma interface.

\subsection{Hycean world scenarios} \label{Res:HyceanOverall}

\begin{figure}
    \centering
    \includegraphics[width=\columnwidth]{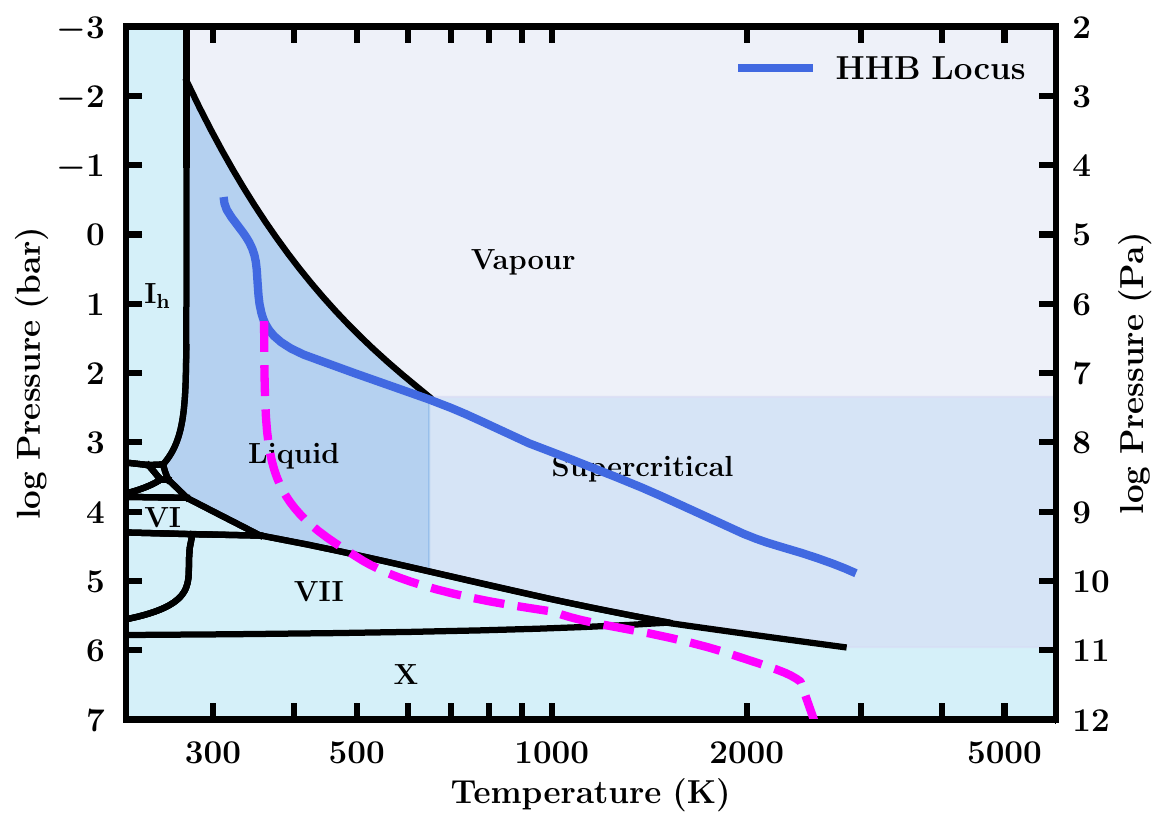}
    \caption{Locus of possible HHB conditions for TOI-270~d for the Case 1 envelope $P$-$T$ profile. An interior adiabat is also shown, corresponding to the hycean cases in Section~\ref{Res:Hycean}, with surface conditions at 360~K and 18~bar.}
    \label{fig:HHBlocus}
\end{figure}

\begin{figure*}
    \centering
    \includegraphics[width=1.9\columnwidth]{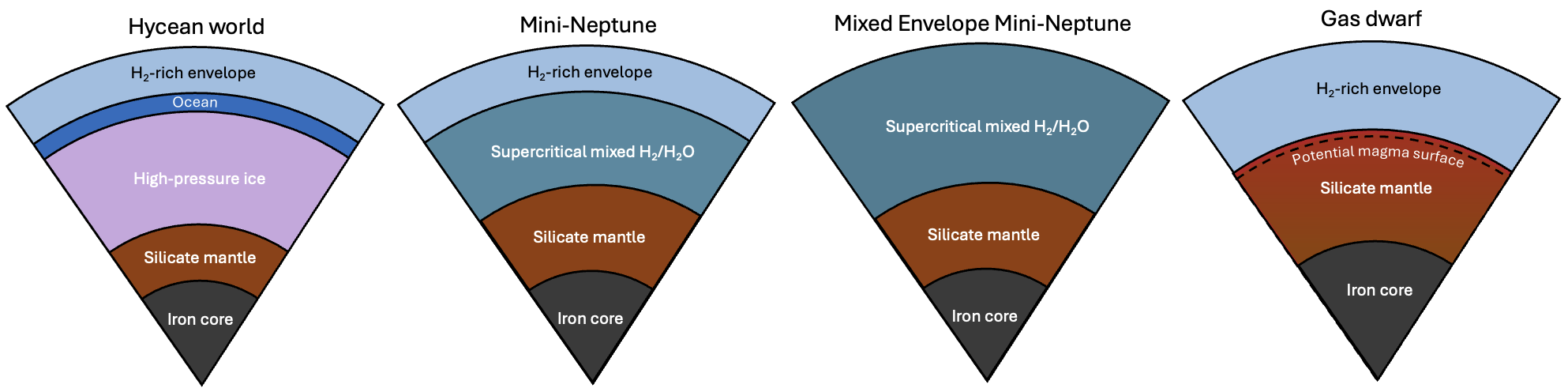}
    \caption{Examples of interior scenarios consistent with the bulk properties of TOI-270~d, spanning different atmospheric temperature structures and water mass fractions. The thickness of the layers are not to scale.}
    \label{fig:cases}
\end{figure*}

We have shown that the pressure and temperature conditions at the surface of TOI-270~d are highly dependent on the assumed temperature structure in the envelope, which in turn defines the phase of water (or silicates, in the gas dwarf scenario) at the surface. The possibility of a liquid H$_2$O surface is permitted by the $P$-$T$ profiles Case 1 and Case 2, with the high haze parameter of $a=1500$ and $T_{\mathrm{int}}=25$ \& $50$~K respectively. As described above, in Figure \ref{fig:HHBlocus} we show the HHB locus for Case 1 that correspond to the solutions in Figure~\ref{fig:ternarycontour}. 

In this section, we explore the solutions that permit a liquid H$_2$O surface on TOI-270~d. For cases assuming efficient day-night heat redistribution, we distinguish between liquid water surfaces that host habitable conditions and those that do not, by the terms "hycean worlds" and "hot hycean worlds". We define a hot hycean world as having a liquid H$_2$O surface beneath an H$_2$-rich atmosphere, with surface conditions beyond the range of commonly considered habitability limits. Habitable conditions are here defined as $273 \leq T_\mathrm{HHB} \leq 395$~K and $1 \leq P_\mathrm{HHB} \leq 1000$~bar, based on conditions where life is found on Earth \citep{Rothschild2001,Merino2019}. We note that the maintenance of a liquid water ocean beneath a H$_2$-rich envelope across the full range of pressures and temperatures used here is debated, due to runaway greenhouse and convective inhibition effects \citep[e.g.][]{Leconte2024} -- this will be discussed further in Section~\ref{Disc:Hycean}. 

TOI-270~d is sufficiently close to its host star that it is likely tidally locked. On tidally locked exoplanets, if the day-night heat redistribution is inefficient, this could lead to large temperature differences between the day and night sides. Dark hycean conditions are defined as a hycean world with a nightside that is sufficiently cool to host habitable conditions while the dayside is too hot. In Section \ref{Res:DarkHycean} we explore possible dark hycean scenarios for TOI-270~d, as suggested as a possibility by \citet{Holmberg2024}.

\subsubsection{Hycean world} \label{Res:Hycean}

We find compositions permitting a liquid H$_2$O ocean with habitable surface conditions, i.e. hycean conditions, are permitted by Cases 1 and 2. We find the range of envelope mass fractions permitting a habitable liquid surface on TOI-270~d to be $\lesssim3.5\times10^{-5}$. The limitation defining the maximum envelope mass is the resulting surface temperature rather than pressure; the Case 1 profile reaches $395$~K at $\sim100$~bar. We find that habitable liquid surfaces require large H$_2$O mass fractions $\gtrsim60\%$ in their interiors. As shown in Figure~\ref{fig:ternarycontour}, the small envelope mass fractions required for low temperature surfaces require a large H$_2$O mass fraction to satisfy the planet's bulk density.

An example hycean interior with median $R_\mathrm{p}$ and $M_\mathrm{p}$ consists of $x_{\mathrm{H_2O}}=75\%$, $x_{\mathrm{env}}=0.0012\%$ and the remainder in an Earth-like core. The HHB in this case is at $T_{\mathrm{HHB}}=360$~K and $P_{\mathrm{HHB}}=18$~bar. The depth of the ocean above the high-pressure ice layer is $284$~km, with the high-pressure ice of thickness $1.08\ \mathrm{R_\oplus}$ (6861~km). The adiabat throughout the H$_2$O layer for this case is shown in Figure~\ref{fig:HHBlocus}.

Previous studies have investigated the ocean depths of water-rich exoplanets, finding a dependence of depth on the surface gravity and surface temperature \citep{Noack2016,Nixon2021,Rigby2024}. For a given surface temperature, a lower surface gravity results in a larger ocean depth. While, for a given surface gravity, a higher surface temperature leads to a deeper ocean for HHB temperatures up to $413$~K, where the trend reverses \citep{Nixon2021} due to the adiabatic temperature structure assumed in the H$_2$O layer. The difference in pressure from the surface to ocean base affects the ocean depth, and the base pressure is fixed by the surface temperature. Varying the surface pressure within the plausible range for a liquid surface has little effect on the ocean depth. The definition of ocean depth adopted by \citet{Nixon2021} is the depth at which the liquid phase ends, either via the transition to high-pressure ice or to supercritical water. For the following, we assume that the HHB pressure is such that the HHB remains in the liquid phase. A surface with $273\leq T_\mathrm{HHB}\lesssim295$~K would result in an ice VI ocean base. However, this is not possible for the envelope temperature structure considered for TOI-270~d. $295 \lesssim T_\mathrm{HHB}\leq413$~K would result in an ice VII base. For $413<T_\mathrm{HHB}<647$~K there would be a supercritical portion of the ``ocean'' between the liquid and high-pressure ice layers, provided -- this is discussed in the following section. 

The contours in the right-hand side of Figure~\ref{fig:ternarycontour} show a kink at $x_\mathrm{env}\sim10^{-4}$ for both core compositions. This corresponds to the HHB at which the H$_2$O adiabat will no longer intersect the ice VII phase boundary with increasing HHB temperature.

We find the possible range of hycean ocean depths to be $\sim215$-$450$~km. For reference, the average depth of Earth's ocean is $3.7$ km \citep{Charette2010} and the deepest point is $\sim$$11$ km \citep{Gardner2014}. The minimum ocean depth is limited by the surface pressure. The depth of $215$~km corresponds to a surface pressure of 1~bar, at a temperature of $337$~K. Conversely, the maximum ocean depth occurs for the maximum habitable surface temperature of $395$~K. This temperature is the maximum considered for habitability, however the runaway greenhouse limit may be reached at temperatures below this \citep[e.g.][]{Leconte2024}. The thickness of the high-pressure ice layers for TOI-270~d with a hycean surface are found to span up to $\sim$$1\mathrm{R_\oplus}$. Thicker high-pressure ice layers correspond to larger H$_2$O mass fractions and lower HHB temperatures. The thickness of high-pressure ice layers may have implications for their habitability \citep[e.g.][]{Madhusudhan2023a}.

\subsubsection{Hot hycean world} \label{Res:HotHycean}

In addition to habitable pressure-temperature surface conditions, we find that interior solutions with a liquid H$_2$O surface at more extreme pressures/temperatures are permitted by both Cases 1 and 2. As shown in Figure~\ref{fig:HHBlocus}, a liquid surface is permitted by Case 1 up until $T_\mathrm{HHB}\leq647$~K, the critical temperature. We note that the runaway greenhouse limit may be reached at temperatures below this \citep[e.g.][]{Leconte2024}, potentially precluding an ocean at these temperatures. For Case 2, a liquid surface is not possible above a HHB temperature of $\sim$500~K where the profile crosses into the vapour region of the water phase diagram, as seen in Figure~\ref{fig:PTh2o}. Thicker atmospheres and hence hotter HHBs will result in a supercritical HHB (or vapour, for Case 2). Under these conditions, it may become important to consider H$_2$O miscible in H$_2$ -- this is considered in Section \ref{Res:MiniNep} for mini-Neptune cases. 

Envelope mass fractions permitting a liquid surface on TOI-270~d, not restricted to habitable pressures and temperatures, are found to be $\lesssim1.5\times10^{-4}$ for the envelope $P$-$T$ profiles considered. The corresponding H$_2$O mass fractions required are found to be $\gtrsim50\%$. An example interior with a liquid surface with the median $R_\mathrm{p}$ and $M_\mathrm{p}$, assuming Case 1 as the envelope $P$-$T$ profile, consists of $x_{\mathrm{H_2O}}=60\%$, $x_{\mathrm{env}}=0.011\%$ and the remainder in an Earth-like core. The HHB in this case is at $T_{\mathrm{HHB}}=584$~K and $P_{\mathrm{HHB}}=172$~bar. The liquid ocean becomes supercritical beyond a depth of $23$~km, down to $1693$~km, before a high-pressure ice layer of thickness $4199$~km (0.66$\mathrm{R_\oplus}$). An example, also with the median $R_\mathrm{p}$ and $M_\mathrm{p}$, instead adopting Case 2 as the envelope $P$-$T$ profile, consists of $x_{\mathrm{H_2O}}=72\%$, $x_{\mathrm{env}}=0.00103\%$ and the remainder in an Earth-like core. The HHB in this case is at $T_{\mathrm{HHB}}=436$~K and $P_{\mathrm{HHB}}=15$~bar. The liquid ocean becomes supercritical beyond a depth of $361$~km, down to $608$~km, and lies atop a high-pressure ice layer of thickness $6331$~km (0.99$\mathrm{R_\oplus}$). To reproduce the same $R_\mathrm{p}$, the hotter envelope $P$-$T$ profile requires a smaller value of $x_{\mathrm{env}}$ for a given value of $x_{\mathrm{H_2O}}$.

As for the habitable surface cases in Section~\ref{Res:Hycean}, we place constraints on the ocean depths permitted for the liquid surface solutions. Overall, we find the possible range of ocean depths for TOI-270~d with a liquid surface to be $\lesssim480$~km. The peak depth occurs at $T_\mathrm{HHB}=413$~K \citep{Nixon2021}, before a supercritical layer beneath the ocean occurs for higher HHB temperatures.

\subsubsection{Dark hycean conditions} \label{Res:DarkHycean}

In this section, we explore the possibility of dark hycean scenarios for TOI-270~d, which we note may be infeasible when considering the effects of atmospheric circulation and the runaway greenhouse limit \citep[e.g.][]{Innes2023,Leconte2024}. It remains difficult to place robust constraints on the day and nightside surface temperatures of tidally-locked sub-Neptunes observed only with transmission spectroscopy, since observations probing the photosphere at the terminator cannot easily constrain the day or night side temperature structure, albedo or atmospheric composition. Therefore, we explore a range of HHB temperatures for the day and night sides of potential dark hyceans. \citet{Innes2023} suggest weak temperature gradients for temperate sub-Neptunes with H$_2$-rich atmospheres due to the slow rotation rates. Informed by this, we therefore consider day-night temperature contrasts up to $100$~K. Our maximum contrast of 100~K is likely an extreme value given the weak temperature gradient suggested by GCM studies of temperate sub-Neptunes \citep[e.g.][]{Innes2023,Benneke2024}. We note that we do not consider the dynamics of the atmosphere or ocean in this initial exploration of the internal structure of dark hycean worlds, which is expected to have a significant effect on the possible day-night contrast. In this highly simplified scenario, we treat the day and night sides independently, adopting an adiabatic profile in the interior on each side for an assumed $T_\mathrm{HHB}$ and $P_\mathrm{HHB}$. For dark hycean conditions, the nightside HHB is at habitable temperatures ($273\leq T_\mathrm{HHB}\leq395$~K) and pressures ($1-1000$ bar). On the dayside we allow the temperature to vary within the full possible range while maintaining a liquid surface -- at $10$~bar this is up to $\sim$$450$~K and at $100$~bar this is up to $\sim$$585$~K. In reality, as discussed above, the day-night contrast would likely be less than this \citep[e.g.][]{Innes2023}. We are also assuming that the hotter day-side surface remains below the runaway greenhouse limit, as potentially possible in the presence of high albedo \citep[e.g.][]{Leconte2024}, as required for a hycean scenario \citep{Madhusudhan2021}. 

\begin{figure*}
    \centering
    \includegraphics[width=0.6\columnwidth]{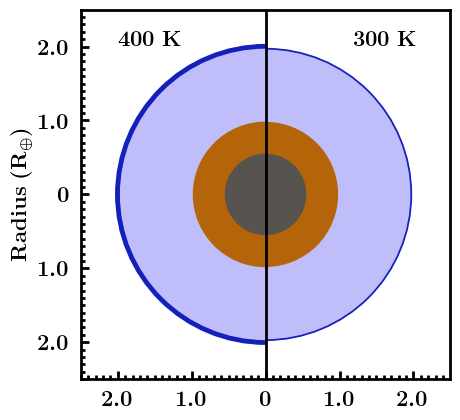}
    \includegraphics[width=0.6\columnwidth]{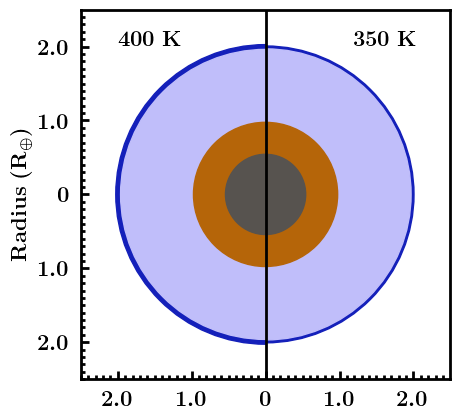}
    \includegraphics[width=0.6\columnwidth]{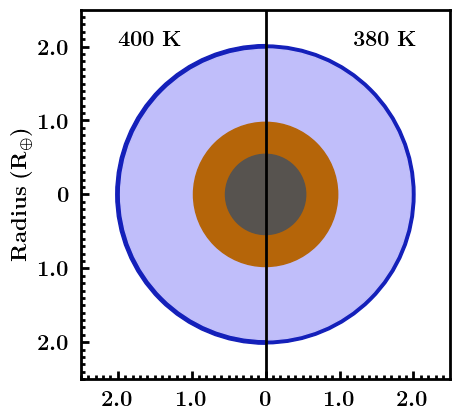}    
    \caption{Examples of possible dark hycean interiors of TOI-270~d, excluding the H$_2$-rich envelope, with varying nightside HHB temperatures. The right-hand-side of each plot shows the nightside with $T_\mathrm{HHB}=300$~K, $T_\mathrm{HHB}=350$~K and $T_\mathrm{HHB}=380$~K, while the left-hand-side of each plot shows the dayside with $T_\mathrm{HHB}=400$~K. All cases have $70\%$ H$_2$O by mass, and $P_\mathrm{HHB}=50$~bar. The dark blue represents the liquid phase of H$_2$O, the light blue indicates high-pressure ices, and the brown and grey represent the mantle and core respectively.}
    \label{fig:darkHyceaninteriorTnight}
\end{figure*}

\begin{figure}
    \centering
    \includegraphics[width=0.6\columnwidth]{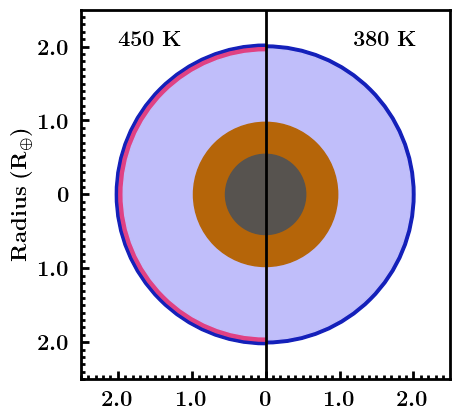}
    \caption{Example of possible dark hycean interiors of TOI-270~d, excluding the H$_2$-rich envelope, with varying day and night side HHB temperatures, with supercritical layers on the dayside. The right-hand-side of this plot shows the nightside while the left-hand-side of this plot shows the dayside. This case has $70\%$ H$_2$O by mass, and $P_\mathrm{HHB}=75$~bar. The dark blue represents the liquid phase of H$_2$O, the pink shows the supercritical region, the light blue indicates high-pressure ices, and the brown and grey represent the mantle and core respectively.}
    \label{fig:darkHyceaninteriorsupercrit}
\end{figure}

In Figure \ref{fig:darkHyceaninteriorTnight} we show examples of dark hycean interiors for TOI-270~d. In these cases, the HHB is fixed at $50$~bar for both the day and night sides, i.e. only the HHB temperature is varying between them. The value of $50$~bar would be the pressure for a $400$~K surface, using the Case 1 $P$-$T$ profile. For these surface conditions, a solution with median $M_\mathrm{p}$ and $R_\mathrm{p}$ has $\sim70\%$ H$_2$O by mass; therefore we adopt this fraction, with the remaining mass in an Earthlike core. We vary the nightside HHB from $300$-$380$~K, for a $400$~K dayside. The corresponding interiors are shown in Figure~\ref{fig:darkHyceaninteriorTnight}. For all of the HHBs shown at $400$~K, $380$~K, $350$~K and $300$~K, the resulting H$_2$O phase structure is liquid to ice VII to ice X. (At this $P_\mathrm{HHB}$, $T_\mathrm{HHB}>413$~K is required for a supercritical portion of the ocean.) We also show an example with the dayside sufficiently hot for a supercritical layer between the ocean and ice, in Figure~\ref{fig:darkHyceaninteriorsupercrit}. The dayside HHB here is at $450$~K and $75$~bar. The effect of ocean dynamics on the existence of supercritical regions is yet to be explored. As is evident from Figures \ref{fig:darkHyceaninteriorTnight} and \ref{fig:darkHyceaninteriorsupercrit}, this difference in surface temperature makes negligible difference to the internal structure beneath the ocean.

\subsection{Mini-Neptune} \label{Res:MiniNep}

For the $P$-$T$ profiles considered in this work, we find a wide range of H$_2$O and envelope mass fractions across which the interior of TOI-270~d could host mini-Neptune-like conditions. These interiors correspond to the vapour or supercritical HHBs shown in Figure \ref{fig:HHBlocus}. An example of a typical mini-Neptune solution consists of 0.4\% H/He envelope, 50\% H$_2$O and 49.6\% Earth-like core. The HHB here is supercritical, at 953~K and~5510 bar, with the H$_2$O transitioning to ice XVIII (superionic ice) deeper in the interior. These calculations assumed that the H$_2$-rich envelope and H$_2$O layers remain distinct. However, it has been suggested that supercritical H$_2$O would become mixed with the H$_2$-rich envelope \citep{Soubiran2015,Gupta2024}. Recently, \citet{Gupta2024} performed molecular dynamics simulations based on density functional theory to establish the interactions of hydrogen and water at pressures and temperatures relevant to sub-Neptune planets. They found that the critical curve, marking the transition from single to separate phases, is at lower temperature than previous studies \citep{Soubiran2015} and can well-match experimental data. Therefore, this underscores the expectation, as suggested by other studies \citep[][]{Nixon2021,Benneke2024}, that a mixed water/hydrogen envelope may be present for sufficiently warm, water-rich sub-Neptunes. \citet{Gupta2024} suggest this to be the case for TOI-270~d, as found by \citet{Benneke2024}, if its temperature at $1$~bar exceeds its zero albedo equilibrium temperature of $387$~K. This is the case for our $P$-$T$ profiles Cases 3 \& 4, while Cases 1 \& 2, with more hazes, are cooler at 1 bar. The median retrieved atmospheric abundances from \citet{Holmberg2024} shows an H$_2$-dominated composition. This would hence correspond to a cold trap, as in Cases 1 \& 2.

We calculate the EOS for varying proportions of H$_2$O by mass in the atmosphere using Equations \ref{eqn:mixeddensity} and \ref{eqn:mixedadgrad}, as in Section \ref{Method:EOSs}. We vary the atmospheric H$_2$O mass proportion ($f_\mathrm{H_2O}$) up to $50\%$ (from our canonical 10\%), somewhat nominally, informed by the uncertainty in the observed H$_2$O mixing ratio. We consider a range of envelope $P$-$T$ profiles by varying the radiative-convective boundary pressure, $P_\mathrm{rc}$. We modify Case 1 to adopt $P_\mathrm{rc}=10$~bar, in addition to the canonical Case 1 profile. Beyond $P_\mathrm{rc}$ the profile follows an adiabat, which we calculate assuming different values of $f_\mathrm{H_2O}$. The envelope composition is considered fully mixed below the cold trap. In all cases, for simplicity, the mass contained in the H$_2$-rich atmosphere above this cold trap is considered to be negligible, and we treat this as isothermal. We assume there is no phase separation throughout the mixed layer, i.e. no formation of high-pressure ices below the envelope \citep{Gupta2024}. The mass and radius are fixed at their median values for the following calculations. 

We first consider the canonical Case 1 profile. For an envelope H$_2$O to H$_2$/He fraction ($f_\mathrm{H_2O}$) of $50\%$ overlying an Earth-like interior we find an envelope mass fraction of $x_{\mathrm{env}}=5.64\%$ is required to achieve the median $R_\mathrm{p}$ for the median $R_\mathrm{p}$. The surface lies at $1.85\times10^5$~bar and $3186$~K. In contrast, from Section~\ref{Res:Interiorcomps}, we showed that our canonical envelope composition ($f_\mathrm{H_2O}$=10$\%$) has the smaller corresponding $x_{\mathrm{env}}=2.25\%$ for an Earth-like interior. For $f_\mathrm{H_2O}=40\%$ and an Earth-like interior we find $x_{\mathrm{env}}=4.12\%$ is required, with the surface at $1.37\times10^5$~bar and $3125$~K. 

Next, we consider Case 1 modified to have $P_\mathrm{rc}$ at 10~bar, which results in a profile hotter than the canonical Case 1. Therefore, for $f_\mathrm{H_2O}=40\%$ and an Earth-like interior, a smaller envelope mass fraction of $x_{\mathrm{env}}=3.32\%$ is required to achieve the median radius for the median planetary mass. The equivalent value for $f_\mathrm{H_2O}$ of $50\%$ is $x_{\mathrm{env}}=5.14\%$.

For the Case 2 profile, we do not vary $P_\mathrm{rc}$ since this is already $\sim$$10$~bar, as seen in Figure~\ref{fig:PTenvelope}. For $f_\mathrm{H_2O}=50\%$ and an Earth-like interior we find $x_{\mathrm{env}}=3.75\%$ is required for median $R_\mathrm{p}$ and $M_\mathrm{p}$. The surface lies at $1.29\times10^5$~bar and $4007$~K. In contrast, our canonical $f_\mathrm{H_2O}=10\%$ H$_2$O envelope composition has the smaller corresponding $x_{\mathrm{env}}=1.10\%$ for an Earth-like interior.

As mentioned in Section \ref{Res:Obs}, Cases 3 and 4 lack a cold trap. This would imply that a mixed H$_2$O/H$_2$ envelope would be likely detectable. For the one offset and simultaneous dual transit case that we adopt for our canonical mixing ratios in Section \ref{Method:PTprofiles}, the upper limit may permit $f_\mathrm{H_2O}$ values of $\sim$40\% by mass in the envelope. However, the observational evidence for H$_2$O is tentative \citep{Holmberg2024}. Future observations may reveal stronger H$_2$O detections that will prompt further investigation into the envelope composition and temperature structure, and the corresponding interior composition, similar to that suggested by \citet{Benneke2024}.

As a demonstration of a fully mixed envelope, we consider the Case 4 profile. For a $f_\mathrm{H_2O}=50\%$ and an Earth-like interior we find $x_{\mathrm{env}}=3.24\%$ is required to achieve the median radius for the median planetary mass. The surface lies at $1.15\times10^5$~bar and $3821$~K. The equivalent for $f_\mathrm{H_2O}=40\%$ is $x_{\mathrm{env}}=2.23\%$, with surface at $7.98\times10^4$~bar and $3668$~K. Our canonical $f_\mathrm{H_2O}=10\%$ case required $x_{\mathrm{env}}=0.86\%$.

\subsection{Gas dwarf} \label{Res:Rocky}

In this section we explore the set of interiors lacking a significant H$_2$O mass fraction, with H$_2$-dominated atmospheres overlying rocky interiors, referred to here as gas dwarfs. Gas dwarf scenarios for temperate sub-Neptunes may have solid or molten surfaces beneath the thick H$_2$-rich envelope. Recently, \citet{Rigby2024b} outlined a framework to establish the feasibility of gas dwarf scenarios for temperate sub-Neptunes, such as TOI-270~d. This study showed that the feasibility of a solid or magma surface is highly dependent on the atmospheric temperature structure, while remaining consistent with internal structure constraints from the planetary bulk parameters. The possibility and observational implications of a potential rocky or magma surface on TOI-270~d will be discussed in Section \ref{Disc:Observations}.

In Section \ref{Res:Interiorcomps} we placed initial constraints on the H$_2$-rich envelope mass fractions permitted by the bulk properties of TOI-270~d. The permitted envelope mass fractions are dependent on the assumed temperature structure in the envelope and the core composition. The maximum envelope mass fraction corresponds to a maximum density pure Fe core with the maximum planet radius, at $x_\mathrm{env}=4.70\%$ for the coolest $P$-$T$ profile (Case 1). With an Earth-like and pure silicate interior, the maximum values of $x_\mathrm{env}$ are $2.25\%$ and $1.44\%$ respectively. In these initial calculations, we did not include an EOS prescription for silicate melt, instead using the EOS from \citet{Seager2007}, as described in Section \ref{Method:EOSs}.

Following from our initial calculations, we now place constraints on the interior compositions allowing a magma or solid rocky surface for our range of $P$-$T$ profiles. We further discuss the possibility of a gas dwarf scenario for TOI-270~d in Section \ref{Disc:Gasdwarf}.

\begin{figure}
    \centering
    \includegraphics[width=\columnwidth]{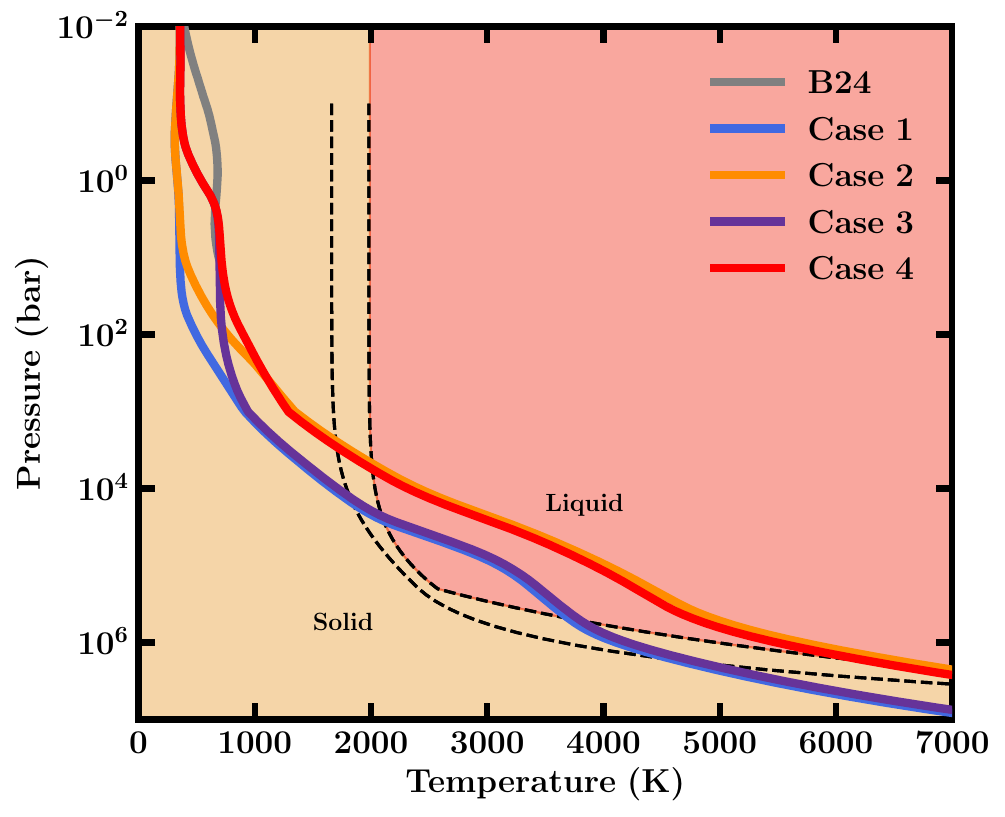}
    \caption{Atmospheric $P$-$T$ profiles considered in this work, compared to the liquidus and solidus for peridotite \citep{Fiquet2010,Monteux2016}. Cases 1-4 are the self-consistent profiles generated in this work from the atmospheric abundances of \citet{Holmberg2024}. We also include the $P$-$T$ profile from \citet{Benneke2024} for comparison.}
    \label{fig:MagmaPT}
\end{figure}

In Figure~\ref{fig:MagmaPT} we show the atmospheric $P$-$T$ profiles Cases 1-4, shown by the solid lines, against the liquidus and solidus for peridotite, shown by the dashed black lines \citep{Fiquet2010,Monteux2016}. Similar to the process for considering H$_2$O-rich interiors, the possibility of a magma surface can be initially assessed by the phase of peridotite at the envelope base. As in \citet{Rigby2024}, the melt-atmosphere interactions and atmospheric chemistry would need to be considered, before establishing the potential feasibility of matching observed atmospheric abundances. We leave an execution of the full framework for TOI-270~d, beyond the internal structure modelling, to future work. 

We first adopt our canonical Case 1 $P$-$T$ profile, the coldest of those considered, and consider interior compositions ranging from the extreme cases of $f_\mathrm{silicate}=100\%$ to $f_\mathrm{silicate}=5\%$. For the following, the solutions have $M_\mathrm{p}$ and $R_\mathrm{p}$ equal to the median measurements. For the extreme case assuming $f_\mathrm{silicate}=100\%$, the surface would not be molten. We note that this is assuming the liquidus as the melt-solid transition -- these pressure/temperature surface conditions lie between the solidus and liquidus and would hence be expected to result in partial melt at the surface. The corresponding $x_\mathrm{env}=1.01\%$ with $P_\mathrm{s}=2.66\times10^4$~bar and $T_\mathrm{s}=2111$~K. With increasing proportion of Fe in the interior, the corresponding surface pressure increases. For an Earth-like interior composition, the surface would be molten, at $P_\mathrm{s}=5.25\times10^4$~bar and $T_\mathrm{s}=2661$~K. The mass fraction contained in the melt ($x_\mathrm{melt}$) is $16.5\%$, and the corresponding envelope mass fraction is $x_\mathrm{env}=1.65\%$. Increasing the Fe content to a Mercury-like interior ($f_\mathrm{silicate}=30\%$) we find $x_\mathrm{env}=2.76\%$ and $x_\mathrm{melt}=10.13$\%, with surface conditions $P_\mathrm{s}=1.18\times10^5$~bar and $T_\mathrm{s}=3210$~K. For the extreme case of $f_\mathrm{silicate}=5\%$, we find a $4.78\%$ melt fraction, with $x_\mathrm{env}=3.85\%$, with surface conditions $P_\mathrm{s}=2.23\times10^5$~bar and $T_\mathrm{s}=3472$~K. 

We also consider Case 2 for the envelope $P$-$T$ profile. This profile is calculated assuming high hazes ($a=1500$) and a $T_\mathrm{int}=50$~K, compared to a $T_\mathrm{int}=25$~K for Case 1. The hotter profile at higher pressures results in a molten surface across all $f_\mathrm{silicate}$ values considered. An example solution for an Earth-like interior composition, has envelope mass fraction $x_\mathrm{env}=0.94\%$. The corresponding surface conditions are $P_\mathrm{s}=3.11\times10^4$~bar and $T_\mathrm{s}=3241$~K, and the $x_\mathrm{melt}$ is $12.25\%$. Due to the hotter $P$-$T$ profile, the surface pressure for the same $f_\mathrm{silicate}$ in the interior is lower for Case 2 than for Case 1.

\section{Summary and Discussion} \label{Discussion}

In this study, we have conducted a theoretical exploration of the range of possible interior compositions and surface conditions for TOI-270~d, based on atmospheric constraints. We placed constraints on the possible compositions that satisfy the reported bulk properties of the planet, adopting self-consistent envelope $P$-$T$ profiles generated using retrieved atmospheric abundances \citep{Holmberg2024}. We presented our results for the theoretically possible interior and surface conditions, which span gas dwarfs, mini-Neptunes and water-rich worlds with liquid ocean surfaces, including hycean worlds, hot hycean worlds and dark hycean worlds.

For our assumed envelope temperature structures, we first explored the possible interior compositions that permit liquid H$_2$O on the surface of TOI-270~d. The H$_2$-rich envelope mass fractions ($x_\mathrm{env}$) across these cases span $\lesssim1.5\times10^{-4}$, while the H$_2$O mass fractions ($x_\mathrm{H_2O}$) are $\gtrsim 50\%$. Habitable hycean conditions at the HHB are possible for envelope mass fractions $\lesssim3.5\times10^{-5}$ and H$_2$O mass fractions $\gtrsim 60\%$, where the surface is defined to be $273\leq T_\mathrm{HHB}\leq 395$~K and $1\leq P_\mathrm{HHB} \leq 1000$~bar. We note that the runaway greenhouse limit may be reached at temperatures below the upper limit of our assumed temperature range \citep[e.g.][]{Leconte2024}. Across the hycean cases, the possible ocean depths are found to span $215-450$~km, atop icy mantles that can exceed thickness of $1\ \mathrm{R_\oplus}$. We additionally explore a simple model of possible dark hycean conditions, where only the nightside has habitable pressures and temperatures at the surface. This was suggested to be a possibility on TOI-270~d due to its relatively high equilibrium temperature for a hycean candidate \citep{Holmberg2024}.

Hycean scenarios are the only shallow-atmosphere interiors compatible with the bulk properties of TOI-270~d. The alternatives are a mini-Neptune and a gas dwarf -- a rocky interior beneath a thick H$_2$ envelope. In Section~\ref{Res:MiniNep} we explored the possibility of a mini-Neptune scenario for TOI-270~d. We adopt a mixed H$_2$O/H$_2$ envelope up to a cold trap, as permitted by the atmospheric $P$-$T$ profiles Cases 1 and 2, which would remain compatible with the retrieved H$_2$O abundance of \citet{Holmberg2024}. For gas dwarf scenarios, considered in Section~\ref{Res:Rocky}, we find the possible envelope mass fractions to span $1-5\%$, with conditions at the surface $\sim10^4-10^5$~bar. Depending on the atmospheric $P$-$T$ profile, we show that magma oceans may be possible on TOI-270~d for these envelope mass fractions. Further study is required to robustly assess the feasibility of a magma ocean surface given the observed atmospheric composition \citep{Rigby2024b,Benneke2024}.

In this section, we explore the possible classification of TOI-270~d and the feasibility of robustly constraining the nature of its surface and interior. We discuss the possibility of TOI-270~d hosting habitable conditions as a hycean or dark hycean world in Section~\ref{Disc:Hycean}. We then discuss the mini-Neptune scenario for TOI-270~d in Section~\ref{Disc:mininep}. In Section~\ref{Disc:Gasdwarf}, we explore the feasibility of a gas dwarf scenario for TOI-270~d, including its potential to host a magma ocean, in addition to the detectability of this compared to a liquid water ocean. Finally, we outline the future prospects for the characterisation of TOI-270~d in light of the recent and upcoming observations.

\subsection{Hycean scenarios} \label{Disc:Hycean}

In Section \ref{Res:HyceanOverall} we investigated interior solutions permitting liquid water on the surface of TOI-270~d. These solutions were explored for our $P$-$T$ profiles Case 1 and Case 2 -- the profiles that pass through the liquid region of the H$_2$O phase diagram. We identified the set of solutions corresponding to habitable hycean scenarios, in addition to liquid surfaces with temperatures $\gtrsim395$~K, which would preclude habitability. We note that the formation pathway of hycean worlds remains unexplored -- for instance, whether the large mass fraction of water can remain as a distinct layer or is at least partially mixed with the rock \citep[e.g.][]{Vazan2022,Luo2024}.

Recent studies of hycean climates have suggested that convective inhibition may result in steep temperature gradients in the atmosphere close to the surface, resulting in surfaces too hot to sustain liquid water \citep[e.g.][]{Innes2023,Leconte2024}. For K2-18~b, which has a lower equilibrium temperature than TOI-270~d, \citet{Leconte2024} found that a high albedo ($A_\mathrm{B}>0.5-0.6$) due to clouds and/or hazes would be required to maintain a liquid ocean. This high albedo is consistent with the definition for hycean worlds given by \citet{Madhusudhan2021}. \citet{Leconte2024} argue that the non-muted CH$_4$ feature in the transmission spectrum of the day-night terminator of K2-18~b may not allow for strong hazes. However, the data does show significant evidence for inhomogeneous clouds/hazes \citep{Madhusudhan2023b} which could provide the required albedo if similar or stronger hazes are present on the dayside. The models of \citet{Innes2023} do not incorporate clouds and hazes. For TOI-270~d, from the atmospheric $P$-$T$ profiles, we have established that hazes are required to allow the possibility of a liquid surface on TOI-270~d -- Cases 1 \& 2 permit liquid oceans at the planetary surface, for a range of internal temperatures, assuming a strong albedo due to hazes. The recent observations did not provide strong constraints on the presence of clouds and/or hazes in the atmosphere of TOI-270~d \citep{Holmberg2024}, which may be refined by future observations.

As suggested by \citet{Holmberg2024}, even if the dayside of TOI-270~d is too warm to allow for habitable conditions, inefficient day-night redistribution could lead to a habitable nightside. In Section \ref{Res:DarkHycean} we explored potential dark hycean scenarios for TOI-270~d. \citet{Innes2023} suggest weak temperature gradients for temperate sub-Neptunes with H$_2$-rich atmospheres due to the slow rotation rates. These processes and resulting temperature contrasts have yet to be established for a moist hycean atmosphere with clouds and/or hazes, therefore we considered a range of surface temperatures, with day-night temperature contrasts up to a high value of $100$~K. The dynamics of hycean atmospheres and oceans have yet to be explored, including the distribution of heat in the oceans in tidally locked hyceans, which will have significant implications for the dark hycean scenario. To maintain a liquid surface, the dayside conditions would further need to remain below the runaway greenhouse limit \citep[e.g.][]{Leconte2024}, requiring a high albedo. 

\subsection{Mini-Neptune scenarios} \label{Disc:mininep}

In Section~\ref{Res:MiniNep} we explored mini-Neptune scenarios for TOI-270~d, which are characterised by a water-rich interior with no distinct surface. We include the possibility of a mixed H$_2$/H$_2$O layer, as has suggested to be likely for sufficiently warm conditions on water-rich sub-Neptunes \citep[e.g.][]{Nixon2021,Benneke2024,Gupta2024}. This situation can arise with or without a cold trap. In the former case, H$_2$O would condense out, resulting in a H$_2$-dominated upper atmosphere, as consistent with the findings of \citet{Holmberg2024}. This is represented in the interior solutions from our $P$-$T$ profiles Cases 1 \& 2. Meanwhile, the results of \citet{Benneke2024}, with high mean molecular weight, are suggested to be consistent with a fully mixed envelope. However, as yet, no strong evidence for H$_2$O has been found for the atmosphere of TOI-270~d \citep[][]{Holmberg2024,Benneke2024,Felix2025}. Future observations are required to strengthen the evidence for H$_2$O, which will provide further insight into the envelope composition and temperature structure.

Recent work has investigated the conditions for miscibility and phase separation of water and hydrogen at pressures relevant to sub-Neptune interiors \citep{Gupta2024}. Future work, including evolution models, should incorporate a more thorough treatment of this behaviour to assess the possible phase behaviour in such interiors. Whether a sub-Neptune's interior temperature structure crosses the critical curve during its evolution, indicating phase separation of hydrogen and water, may prove crucial to assessing its present interior structure. 

The presence of abundant NH$_3$ and CH$_4$ have been suggested as one characteristic of a mini-Neptune atmosphere, due to the deep atmosphere recycling of photochemically produced nitrogen and carbon species \citep[e.g.][]{Tsai2021,Hu2021,Cooke2024}. The atmospheric chemistry of TOI-270~d and similar sub-Neptunes will vary depending on the presence or lack of a cold trap \citep[e.g.][]{Yang2024}. Future observations providing improved constraints on the H$_2$O abundance will be crucial to establish this. Further work is additionally required to strengthen the observational predictions for a mini-Neptune scenario, including accurate and complete cross-section data, and chemical networks for photochemical models. We also note that the chemistry in a supercritical mixed H$_2$/H$_2$O envelope has not been explored in detail.

\subsection{Gas dwarf scenarios} \label{Disc:Gasdwarf}

The set of compositions permitting a gas dwarf scenario for TOI-270~d, characterised by a rocky interior and a thick H$_2$-rich envelope, were explored in Section \ref{Res:Rocky}. For our assumed atmospheric temperature profiles, we find the envelope mass fractions required in this scenario to be $\sim1-5\%$, for interior compositions varying between the extreme cases of pure silicate to pure iron interiors. The corresponding atmospheric base pressures span $\sim10^4-10^5$~bar, depending on the assumed core composition.

Recently, \citet{Rigby2024b} outlined a method for evaluating the plausibility of gas dwarf scenarios, including the presence of a magma ocean, for temperate sub-Neptunes. This end-to-end framework coupled an internal structure model, melt-atmosphere interface chemistry and atmospheric models. This study found that a high CO/CO$_2$ ratio ($>>1$) is a key indicator of a deep atmosphere. Taking into account the bulk properties of the planet in a holistic modelling framework is hence crucial to establish the feasibility of a gas dwarf scenario and potential presence of a magma ocean. They hence find using this framework that the observations for K2-18~b could not be explained by a magma ocean. The pattern of detections and non-detections in TOI-270~d is similar to those of K2-18~b \citep{Madhusudhan2023b}, including the depletion of NH$_3$ \citep{Holmberg2024,Benneke2024,Felix2025}. Although, similarly to K2-18~b, JWST observations of TOI-270~d thus far have provided only upper limits on the CO abundance \citep{Holmberg2024,Benneke2024}. Given these upper limits, a CO/CO$_2$ ratio $>>1$ is unlikely for TOI-270~d, which would be difficult to reconcile with the presence of a deep, gas dwarf atmosphere. \citet{Rigby2024b} also tentatively suggest that sulfur depletion could be an indicator, depending on the assumed solubility laws, which could become important for TOI-270~d given the tentative CS$_2$ detection. Recently, \citet{Glein2025} suggested a lack of CO in the atmosphere of TOI-270~d could be explained by equilibrium chemistry of hot gas in a H$_2$O-rich and CO$_2$-poor atmosphere. This study relies on the presence of a high mean molecular weight atmosphere \citep{Benneke2024,Felix2025}, which is in contrast to the findings of \citet{Holmberg2024}.

For TOI-270~d, as shown in this work, and K2-18~b as shown in \citet{Madhusudhan2020}, large atmospheric mass fractions $\gtrsim1\%$ are required to satisfy the observed mass and radius in the absence of a H$_2$O layer. Under these conditions, surface pressures at the lowest are $\sim$$10^4$~bar and for many interior solutions for both planets extend to $\gtrsim10^5$~bar \citep{Madhusudhan2020,Rigby2024}. The partitioning of volatiles at high pressures ($\gtrsim$$10^{5}$~bar) is not well tested experimentally \citep{Kite2019,Schlichting2022,Shorttle2024}, therefore future studies are needed to further explore the interaction of a magma ocean with a H$_2$-rich atmosphere at such extreme pressures and temperatures. Modelling gas dwarfs with potential magma oceans also requires an assumption to be made on the nature of the melt composition, which can have significant effects on the density and melt curve. Furthermore, the phase behaviour of hydrogen and rock at such high pressures \citep[e.g.][]{Young2024} may have implications for the differentiation of these materials, and hence the formation of a distinct surface.

Establishing the nature of planets in the sub-Neptune regime is crucial in understanding planet formation and evolution processes, including the cause of the radius valley \citep[e.g.][]{Fulton2017,Fulton2018,Cloutier2020}. Rocky planets with thick H$_2$-rich envelopes have been suggested to dominate the larger radius peak above the radius valley, with the smaller radius population having lost their H$_2$-rich atmospheres via photoevaporative \citep[e.g.][]{Lopez2013,Jin2014,Owen2017,Jin2018} and/or core-powered mass loss \citep[e.g.][]{Ginzburg2018,Gupta2019,Gupta2020}. However, this remains debated -- for instance, whether water-rich sub-Neptunes also explain this second peak \citep[e.g.][]{Zeng2019,Rogers2025}. For individual exoplanets, atmospheric molecular abundances, in addition to improved precision of mass and radius measurements, are crucial for characterising their interiors and surfaces. As discussed above, further theoretical and experimental studies are also necessary to achieve this. Multiple sub-Neptunes have already been observed with JWST \citep[e.g.][]{Madhusudhan2023b,Kempton2023,Wallack2024}, however to better understand the nature and diversity of this population, high-precision transmission spectra will need to be obtained for large numbers of these planets.

\subsection{Prospects for future characterisation} \label{Disc:Observations}

TOI-270~d has been observed as part of multiple JWST programs (GTO Program 2759, GO Programs 3557 \& 4098), with the NIRISS, NIRSpec and MIRI instruments, spanning 0.9-12 $\mu$m. \citet{Holmberg2024} analysed recent NIRSpec G395H observations in addition to HST data \citep{MikalEvans2023}, reporting strong detections of CH$_4$ and CO$_2$ detections and a non-detection of NH$_3$, along with tentative H$_2$O and CS$_2$. As we have shown in this work, the planetary bulk and atmospheric properties allow for a wide range of interior compositions for TOI-270~d, given the current uncertainties.

Constraints on the photospheric metallicity and temperature through transmission spectroscopy can provide useful information on the possible interior compositions. As discussed by \citet{Benneke2024}, their high obtained atmospheric metallicity and temperature would likely preclude the presence of a distinct H$_2$O layer, as this would be supercritical at depth, and hence mixed with H$_2$. The median retrieved abundances of \citet{Holmberg2024} are not compatible with a fully mixed envelope without a cold trap. If future observations were to provide stronger evidence for H$_2$O at higher abundance, towards the upper limit of the current \citet{Holmberg2024} error bars, this may indicate the presence of a mixed H$_2$O/H$_2$ envelope. Robust atmospheric retrievals on precise JWST data are essential to accurately inform these inferences for the interior and surface conditions of sub-Neptunes, including TOI-270~d.

It is difficult to place robust constraints on the day and nightside surface temperatures of tidally-locked sub-Neptunes observed with transmission spectroscopy alone. Atmospheric abundances in the photosphere at the terminator cannot easily constrain the day or night side temperature structure, albedo or composition of the atmosphere. Determining the atmospheric temperature structures of sub-Neptunes is already complex \citep[e.g.][]{Piette2020,Innes2023,Leconte2024}, for instance, due to the effects of clouds/hazes. Therefore, to place constraints on the dayside properties of tidally-locked sub-Neptunes, future observations will need to include emission spectroscopy. 

The range of explanations for the transmission spectra of the recently observed sub-Neptunes highlights the importance of studies into the complex relationship between sub-Neptune atmospheres and surfaces \citep[e.g.][]{Hu2021,Rigby2024b,Yang2024,Cooke2024,Young2024,Glein2025}. The abundance of high precision spectroscopic data from JWST requires robust theoretical models to best inform interpretation of such results. As discussed earlier in Section \ref{Discussion}, additional experimental and theoretical work is needed to establish the observational diagnostics of possible surface layers, including liquid water and magma oceans. Further experimental work is required to improve the solubility data for chemical species in these materials, and to investigate the behaviour of planetary materials at high pressures and temperatures \citep[e.g.][]{Young2024,Gupta2024}. Atmospheric data remains crucial for helping to break degeneracies in interior composition that are inherent in internal structure modelling. Robust data reduction and retrieval frameworks must be employed to accurately determine atmospheric abundances from the high-precision JWST data.

Temperate sub-Neptunes orbiting M dwarfs, including TOI-270~d, remain exciting targets for detailed atmospheric characterisation with JWST. Observations of these planets provide important case studies for understanding the broader nature of the sub-Neptune population and the implications for planet formation/evolution processes. Robustly determining the nature of sub-Neptune atmospheres, surfaces and interiors requires further theoretical and experimental studies, including atmosphere/surface interactions, and equations of state at relevant pressures and temperatures. The high precision transmission spectra of sub-Neptunes, including TOI-270~d, from JWST will help to shed light on their individual compositions and overall on this intriguing planetary regime.

\section*{Acknowledgements}

F.E.R. and N.M. acknowledge support from the Science \& Technologies Facilities Council (STFC) towards the PhD studies of F.E.R. (UKRI grant 2605554). We thank the anonymous reviewer for their comments which improved the manuscript.

\section*{Data Availability}

This work is theoretical and hence no new data is generated.
 


\bibliographystyle{mnras}
\bibliography{Sources} 







\bsp	
\label{lastpage}
\end{document}